\documentclass[lettersize,journal]{IEEEtran}
\usepackage{amsmath,amsfonts,bm}
\usepackage{url}
\usepackage{verbatim}
\usepackage{graphicx}
\usepackage{cite}
\usepackage{hyperref}
\usepackage{subcaption}
\usepackage{enumitem}
\usepackage{multirow}
\usepackage{wrapfig}

\begin{document}

\title{Deep Polarimetric HDR Reconstruction}

\author{Juiwen Ting, Moein Shakeri, Hong Zhang \thanks{The authors are with Department of Computing Science, University of Alberta, Canada}
\thanks{Manuscript received April 19, 2021; revised August 16, 2021.}}

\markboth{Journal of \LaTeX\ Class Files,~Vol.~14, No.~8, August~2021}%
{Shell \MakeLowercase{\textit{et al.}}: A Sample Article Using IEEEtran.cls for IEEE Journals}


\maketitle

\begin{abstract}
This paper proposes a novel learning based high-dynamic-range (HDR) reconstruction method using a polarization camera. We utilize a previous observation that polarization filters with different orientations can attenuate natural light differently, and we treat the multiple images acquired by the polarization camera as a set acquired under different exposure times, to introduce the development of solutions for the HDR reconstruction problem. We propose a deep HDR reconstruction framework with a feature masking mechanism that uses polarimetric cues available from the polarization camera, called Deep Polarimetric HDR Reconstruction (DPHR). The proposed DPHR obtains polarimetric information to propagate valid features through the network more effectively to regress the missing pixels. We demonstrate through both qualitative and quantitative evaluations that the proposed DPHR performs favorably than state-of-the-art HDR reconstruction algorithms. 
\end{abstract}


\section{Introduction} \label{sec:intro}
\IEEEPARstart{I}{n} real-world scenes, the luminance varies over several orders of magnitude and is presented as high-dynamic-range (HDR). Unfortunately, standard digital cameras can only capture a limited fraction of this range due to sensor constraints. As a result, the images captured by such cameras are low-dynamic-range (LDR) with over- or under-exposed regions that do not reflect the human visual system's ability to simultaneously identify details in both bright and dark regions of a scene. In LDR images, information is lost in the over- or under-exposed regions, and thus presents challenges to computer vision related robotic tasks \cite{Shakeri_2016_IROS, Shakeri_2017_ICCV, Shakeri_2019_CVPR}. In HDR reconstruction, the method seeks to restore the lost information, and aims to output an HDR image that prevails a wider range of illumination than is available directly from the camera sensor.

Classical HDR reconstruction methods propose to generate an HDR image from a sequence of LDR images of the scene taken at different exposures \cite{Yan_AttnHDR, Lee_ExpBlendHDR}. However, these approaches require multiple exposure images as input, which may not always be available and practical. Single-shot HDR techniques resolves the above issue by requiring only a single capture. As a result, they have attracted considerable research efforts in recent years.

\begin{figure}[t!]
\centering
\begin{subfigure}{.4\textwidth}
  \centering
  \includegraphics[width=0.95\linewidth]{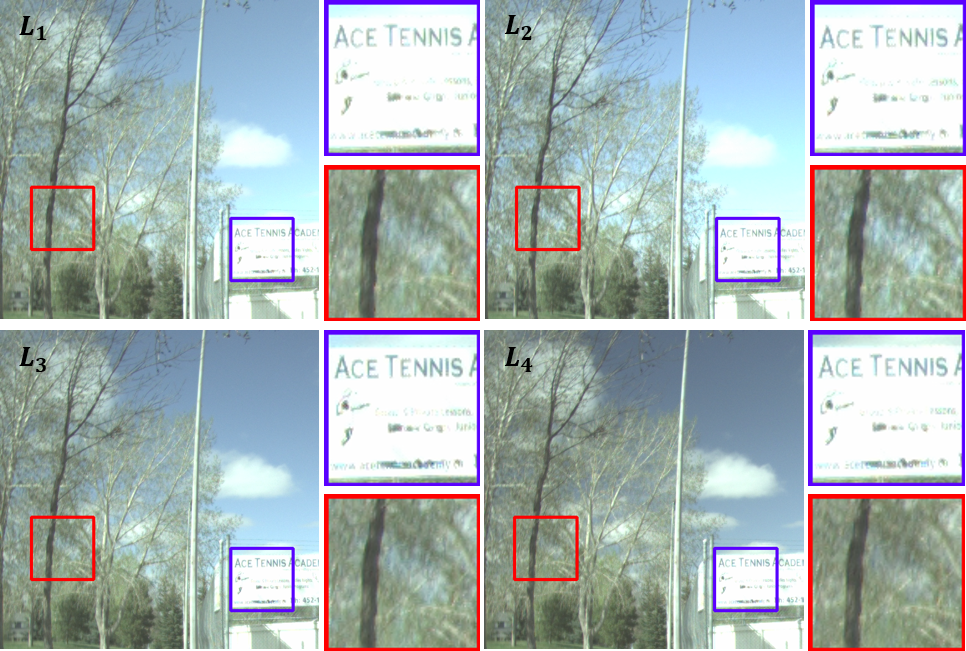}
  \caption{Input LDRs (we show the other input $M_{1,i}$ in Fig.~{\ref{fig:DSSHDRNet}})}
  \label{fig:intro_img_input}
\end{subfigure}%
\centering
\vskip 5pt
\begin{subfigure}{.192\textwidth}
  \centering
  \includegraphics[width=0.983\linewidth]{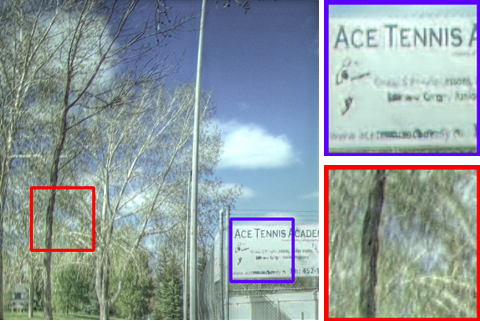}
  \caption{DPHR (final result)}
  \label{fig:intro_img_ours}
\end{subfigure}%
\begin{subfigure}{.192\textwidth}
  \centering
  \includegraphics[width=0.983\linewidth]{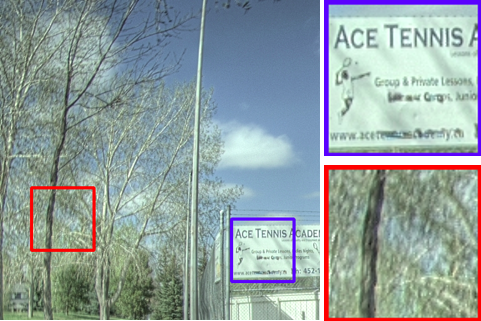}
  \caption{ground truth}
  \label{fig:intro_img_gt}
\end{subfigure}%
\caption{Sample HDR reconstruction result of the proposed DPHR method.}
\label{fig:intro_img}
\end{figure}

A plethora of single-shot HDR works have been proposed \cite{Marnerides, Eilertsen, Santos, Endo, Lee_2018a, Liu}. However, almost all existing methods have some significant drawbacks including ignoring reconstruction for under-exposed regions, neglecting dynamic range expansion for properly exposed pixels, or generating artifacts (i.e., color distortions, halo artifacts). An interesting recent development of imaging technology is a polarization camera that uses multi-directional on-chip linear polarizers to capture multi-polarization images in single-shot. Polarization is a property of light, which can convey rich illumination information about the environment to help with HDR reconstruction. Recent studies show that the polarization camera can handle the aforementioned drawbacks due to the acquisition of the different polarization filtered images. These polarization filtered images can capture different intensity values, where a region that is poorly exposed in one image may be properly exposed in another image. An existing polarimetric method by Wu et al. \cite{xuesong} is a traditional model based approach, and is vulnerable to deal with a) poorly exposed pixels in polarized images, and b) pixels with low degree of polarization. Another existing polarimetric method by Ting et al. \cite{ting2021} is a deep-learning-based approach, whose input consists of the four images from the polarization camera. However, where the degree of polarization is high, the method proposed by Ting et al. \cite{ting2021} is unable to take advantage of the known physics of the polarization images as compared to the method by Wu et al. \cite{xuesong}. Specifically, Ting et al. \cite{ting2021} does not utilize degree of polarization to construct HDR images.

Motivated by these shortcomings, we present a deep-learning-based solution to the problem of HDR reconstruction using a polarization camera, called Deep Polarimetric HDR Reconstruction (DPHR). We design our approach based on two main observations. First, pixels in an LDR image can be a) under-exposed, b) properly exposed, or c) over-exposed. In HDR imaging, the goal is to not only recover the lost information in poorly exposed regions, but to also increase the bit-depth of the properly exposed pixels beyond 8-bits. Therefore, we use the polarimetric cues to design a novel reconstruction process to a) identify reconstruction regions based on our feature masking mechanism for the convolutional neural network (CNN) to focus its learning upon, and b) propagate the properly exposed pixels that contains the valid features through the network. Second, the method by Wu et al. \cite{xuesong} is able to expand the dynamic range in areas that have a strong measure of polarized light. Therefore, we combine the HDR images generated by a traditional model based approach \cite{xuesong} and our deep-learning-based DPHR model to obtain the final HDR result. We evaluate the effectiveness of our method on outdoor scenes captured by a polarization camera. Fig.~\ref{fig:intro_img} shows a sample result of the proposed DPHR method. The main contributions of our work are as follows:

\begin{itemize}
\item We propose a feature masking mechanism to solve the HDR reconstruction problem using a CNN and a polarization camera. This masking approach is based on the domain knowledge of polarization, and effectively identifies the invalid content that requires HDR recovery. Thus, the quality of the final results expands the dynamic range by synthesizing realistic textures and has reduced artifacts.

\item We introduce an HDR reconstruction formulation that combines the complementary results from the traditional model-based and deep-learning-based approaches to generate the final output.
\end{itemize}

The remainder of this paper is organized as follows. Related works on HDR reconstruction and the background physics of light polarization are summarized in Section \hyperref[sec:RelatedWorks]{2}. Section \hyperref[sec:ProposedMethod]{3} presents the details of our proposed DPHR method based on the polarization camera. Experimental results and discussions are described in Section \hyperref[sec:Experiments]{4}, and concluding remarks in Section \hyperref[sec:Conclusion]{5}.


\section{Related Works} \label{sec:RelatedWorks}
In this section, we first review the relevant works to the problem of single-shot and multiple-shot HDR reconstruction, and then explain the background physics of light polarization that is the basis of our proposed method.

\subsection{Multiple-shot HDR Reconstruction} \label{ssec:multiple-shot_HDR_reconstruction}
Multiple-shot HDR reconstruction is a well-known area of research with a series of works that focuses on removing ghosting artifacts caused by moving objects or misalignment in the images shot from multiple acquisitions. Several traditional works perform alignment and HDR reconstruction in a unified optimization system. For example, Sen et al. \cite{Sen12} and Hu et al. \cite{Hu13} proposed patch-based optimization systems to fill in the missing details due to the saturated regions in the reference image using other images within the stack. Other works by Oh et al. \cite{Oh15}, Lee et al. \cite{Lee14} and Li et al. \cite{Li14} use rank minimization where image misalignments are considered as sparse outliers, to align and reconstruct an HDR image. However, the above techniques are not data-driven and produce unsatisfactory results in challenging cases where the reference has pronounced saturated regions.

To alleviate the aforementioned issues, a group of works explores learning based approaches to directly align and reconstruct pixels in all saturated regions. Some works proposed a pre-processing step to align the multi-exposure images prior to feeding them to a CNN. For example, pre-processed steps based on optical flow and homography transformation are proposed, respectively by Kalantari et al. \cite{Khademi_2017} and Wu et al. \cite{wu2018deep}. Other works implemented an end-to-end HDR framework. For example, Yan et al. \cite{Yan_AttnHDR}, Liu et al. \cite{liu2021adnet} and Pu et al. \cite{Pu_2020_ACCV} presented attention-merging networks to generate ghost-free HDR imaging. Ma et al. \cite{ma_mef} proposed an attention-merging network with guided image filtering to optimize the fusion weight maps of the multi-exposure images. Niu et al. \cite{niu2020hdrgan} implemented a generative adversarial network (GAN) based model to fuse the multi-exposure images into an HDR image. In a similar vein, Kalantari et al. \cite{Kalantari2019deepvideo} uses two connected networks to fuse multiple LDR frames, and generate an HDR video. All of these methods need multiple shots, though some methods have accounted for motions between the input images \cite{Yan_AttnHDR, liu2021adnet, Pu_2020_ACCV}, suppressing severe motions remain a challenge thus is not practical in applications with moving cameras. On the other hand, perfectly aligned polarization images can be captured in just a single-shot, which is suitable for applications with both stationary and moving cameras.


\subsection{Single-shot HDR Reconstruction} \label{ssec:single-shot_HDR_reconstruction}
Single-shot HDR reconstruction is a well-studied area of research with a plethora of solutions developed for various applications in computer vision and robotics. A classical and important group of these solutions is expansion operators, where different heuristic operators are implemented to transform a single LDR image to an HDR image\cite{AFR07, RTS07, MSG17, KO14, Masia16, banterle2006inverse, kasliwal2015tonemap}. However, these techniques make assumptions about the scene and use heuristics that result in deviation from the true luminance value, and thus tend to induce pixel artifacts.

To address these limitations, the second group of methods take advantage of the power of CNNs and directly expands the dynamic range to recover pixels in poorly exposed regions. Several novel architectures have been proposed to utilize contextual information extracted by convolutional layers to improve HDR reconstruction results \cite{Marnerides, Eilertsen, Santos, Endo, Lee_2018a, Liu}. For example, Marnerides et al. \cite{Marnerides} proposed a three-branch CNN to extract global, semi-local, and local features that exploit feature recovery at various scales, for reconstruction in all poorly exposed regions. However, their method outputs overly bright and smooth images as it over-enhances the extracted illumination features. Liu et al. \cite{Liu} presented an HDR framework that incorporates the domain knowledge of the LDR image formation pipeline. They trained three specialized networks to reverse the image formation steps of dynamic range clipping, non-linear mapping from a CRF, and quantization to reconstruct an HDR image. Chen et al. \cite{chen2021hdrunet} proposed a U-Net like network to solve the joint tasks of HDR reconstruction, denoising, and dequantization. Zhang et al. \cite{Zhang2021DeepHE} designed a three-stage cascade network to decompose the LDR image to learn the HDR content. Eilersten et al. \cite{Eilertsen} proposed a U-Net like architecture to regress details in over-exposed regions with a fixed mask, and later combined the prediction with the input image for the final result. Santos et al. \cite{Santos} developed an autoencoder structure with a feature masking mechanism to also regress details in over-exposed regions, and then reconstruct the HDR image through combining with the input image for the end result. Namely, the feature masking mechanism by Santos et al. \cite{Santos} is based on a soft mask computed from the pixel intensity of the input image. Thus, dark pixels in the soft mask correspond to over-exposed pixels where the CNN focuses its learning upon, and the soft mask is continuously updated by the feature weights of each convolutional layer. However, both methods by Eilersten et al. \cite{Eilertsen} and Santos et al. \cite{Santos} neglect HDR reconstruction for under-exposed regions, and is unable to improve HDR estimation for properly exposed regions, as it applies a mask that focuses the reconstruction on over-exposed regions. Furthermore, the methods by Eilersten et al. \cite{Eilertsen} and Santos et al. \cite{Santos} has unreliable reconstruction in all poorly exposed regions due to the lack of prior to enforce consistency.

The third group of methods rely on specialized cameras or filters, such as neuromorphic camera \cite{Han2020neuromorphic}, saturation self-reset camera \cite{Bardallo2016AsyncCamera}, modulo camera \cite{Zhao2015ModuloCamera}, snapshot camera \cite{Cull2015SnapshotCamera, Nayar, Suda}, polarization camera \cite{xuesong} or cross-screen filter \cite{Rouf2011Glare} to recover HDR images. Alternatively, some works presented end-to-end joint learning of optics and HDR reconstruction. For example, Metzler et al. \cite{metzler2019deep} and Sun et al. \cite{Sun_2020_LearnedOpticHDR} jointly trained an optical-based element and a CNN to hallucinate the HDR content from a single LDR image. As another example, Martel et al. \cite{Martel2020NeuralSensor}, Alghamid et al. \cite{Alghamdi2021Snapshot, Schulz2019CodeMask}, Fotiadou et al. \cite{Fotiadou2020Snapshot} and Serrano et al. \cite{Serrano2016Snapshot} proposed joint designs to optimize a snapshot sensor and a deep learning algorithm to reconstruct HDR images A snapshot camera is an imaging device capable of acquiring multiple images in a single capture. This is beneficial for HDR imaging as more information (i.e., luminance) about the scene can be captured in a single-shot, to help recover the scene's dynamic range. For example, Nayar et al. \cite{Nayar} proposed a learning based approach to handle the images captured by a spatially varying exposure camera (i.e., multiple exposure images), and generate an HDR image. Recently, traditional model based \cite{xuesong} and deep-learning-based \cite{ting2021} approaches using a polarization camera have been proposed for HDR reconstruction, and showed promising results. However, the method by Wu et al. \cite{xuesong} is vulnerable to challenging cases with significant poorly exposed and low degree of polarization pixels. Though the method by Ting et al. \cite{ting2021} can reconstruct better details than Wu et al. \cite{xuesong}, Ting et al. \cite{ting2021} still outputs limited details. In contrast to the method by Ting et al. \cite{ting2021} that uses only the polarization images to achieve HDR reconstruction, we propose an enhanced deep-learning-based HDR reconstruction method called DPHR, that uses both the polarization images and polarimetric cues captured by the polarization camera to reconstruct HDR images as will be explained in Section \hyperref[sec:ProposedMethod]{3}.

\begin{figure*}[t!]
\centering
\includegraphics[width=0.95\textwidth]{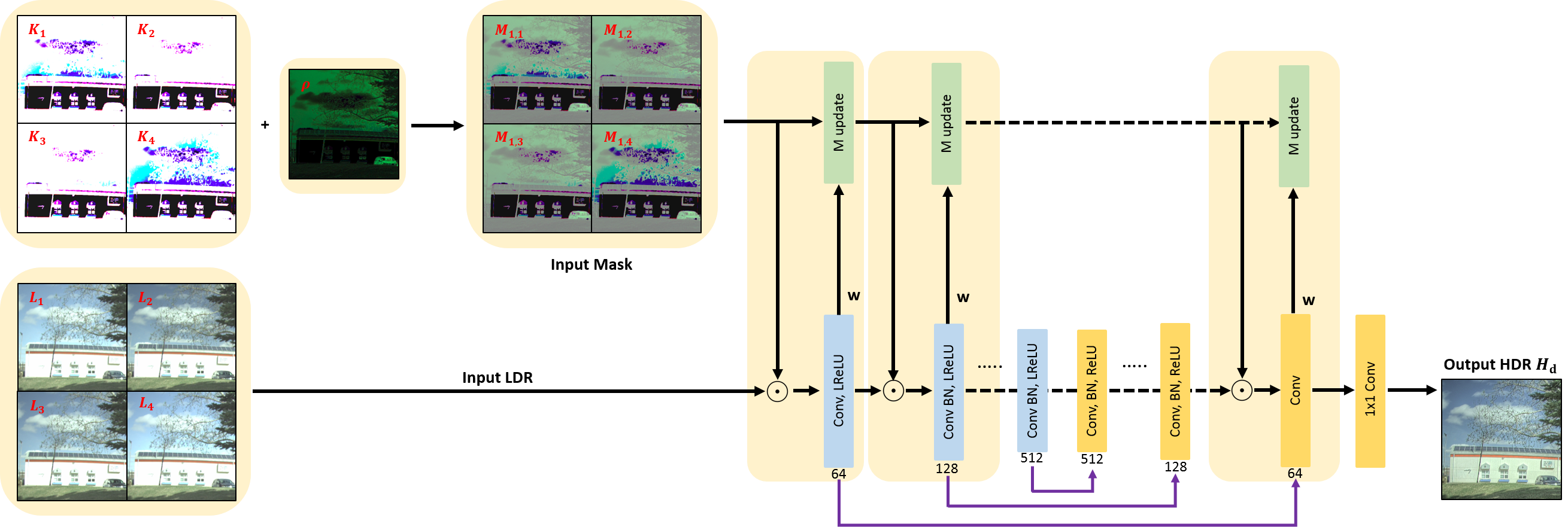}
\caption{Overview of the deep neural network used in constructing $H_d$ in our proposed DPHR framework. The LDR input images $L_i$, $i$ = 1,2,3,4 are propagated through the network, while updated by the corresponding mask $M_{l,i}$ before going through the convolutional layers. The mask at each layer is obtained by updating the mask from the previous layer.}
\label{fig:DSSHDRNet}
\end{figure*}

\subsection{Background Physics of Light Polarization} \label{ssec:polarization_theory}
The proposed DPHR method is based on a polarization camera, which implements pixel-level polarizer filters that has made it possible to acquire real-time and high resolution measurement of incident polarization information. Each calculation unit has four pixels with four on-chip directional polarizer filters, at 0$^\circ$, 45$^\circ$, 90$^\circ$, and 135$^\circ$, to capture four spatially aligned polarization images. Each polarization image is a filtered version of the image irradiance $I$ captured by the polarization camera, and thus the effect of a polarizer on image irradiance can be written as \cite{Stokes}:
\begin{equation}
\begin{gathered}
I_0, I_{90} = 0.5 \times I(1 \pm \rho\cos2\theta)\\
I_{45}, I_{135} = 0.5 \times I(1 \pm \rho\sin2\theta)
\label{eqn:polarizer_irradiance}
\end{gathered}
\end{equation}

\noindent where $\rho \in$ [0,1] denotes the degree of linear polarization (DoLP), and measures the portion of light that is polarized for a given pixel. $\theta \in$ [0$^{\circ}$, 180$^{\circ}$] denotes the angle of linear polarization (AoLP), and measures the direction of light in which the polarized light oscillates at a given pixel. The DoLP can be defined as follows: 
\begin{equation}
\rho = \frac{I_{max} - I_{min}}{I_{max} + I_{min}}
\label{eqn:dop}
\end{equation}

\noindent where $I_{max}$ and $I_{min}$ are the maximum and the minimum measured radiance at every pixel, respectively. These two parameters along with AoLP $\theta$ can be determined by three or more polarization images with different but known polarizer filter angles. As our camera can simultaneously capture four images, the DoLP and AoLP can be readily computed, and their closed form solutions can be found in the work by Huynh et al. \cite{Huynh_2010}.

In general, the relation between image irradiance $I_i$ and pixel value $L_i$ at exposure time $t_0$ can be written as \cite{Debevec}:
\begin{equation}
L_i = f(I_{i}t_0)
\label{eqn:irradiance_pixel}
\end{equation}

\noindent where $f$ is the camera response function. By substituting (\ref{eqn:polarizer_irradiance}) into (\ref{eqn:irradiance_pixel}), and applying the reciprocity relation in the work by Debevec and Malik \cite{Debevec}, we obtain:
\begin{equation}
\begin{gathered}
t_1, t_3 = 0.5 \times t_0 (1 \pm \rho\cos2\theta)\\
t_2, t_4 = 0.5 \times t_0 (1 \pm \rho\sin2\theta)
\label{eqn:HDR}
\end{gathered}
\end{equation}

The polarization image formation model in the case of HDR reconstruction is provided by (\ref{eqn:HDR}). The key observation is that when the incoming light is not entirely unpolarized ($\rho \neq$ 0), the four polarizer orientation images experience different exposure times, effectively creating the condition for multiple exposures. In other words, each polarizer orientation image covers a different luminance range, and thus can be combined to recover a wider dynamic range by synthesizing realistic textures.


\section{Proposed Method} \label{sec:ProposedMethod}
Fig.~\ref{fig:DSSHDRNet} shows an overview of the deep neural network used in constructing $H_d$ in our proposed DPHR framework. The DPHR network architecture is summarized in Table~\ref{tab:chapt4_DSSHDR_network_config}. Our goal is to reconstruct HDR values for all properly exposed and poorly exposed pixels. We achieve this using a CNN that takes as input four LDR polarization images captured by the polarization camera, and estimates the missing HDR information in all properly exposed and poorly exposed regions with the help from the feature masking mechanism. We compute an input mask image for the feature masking mechanism. Then the input mask updates the features at each convolutional layer using the obtained polarimetric information, in order to reduce the magnitude of the features generated by the invalid areas, and thus better propagate the valid features. Finally, we combine the predicted HDR image with the HDR image generated from the traditional model based approach \cite{xuesong} to obtain the final HDR result.

\begin{table}[t!]
\caption{Overview of DPHR network architecture}
\begin{center}
\label{tab:chapt4_DSSHDR_network_config}
\setlength\tabcolsep{1.1pt} 
\begin{tabular}{cccccc}
\hline
\small{Layer} & \small{Stage} & \small{\# Filters} & \small{Filter size} &\small{Conv. stride} &\small{Activation}\\ 
\hline
\small{1} & \small{conv+act} & 64 & (7,7) & (2,2) & \small{ReLU}\\
\hline
\small{2} & \small{conv+bn+act} & 128 & (5,5) & (2,2) & \small{LeakyReLU}\\
\hline
\small{3} & \small{conv+bn+act} & 256 & (5,5) & (2,2) & \small{LeakyReLU}\\
\hline
\small{4} & \small{conv+bn+act} & 512 & (3,3) & (2,2) & \small{LeakyReLU}\\
\hline
\small{5} & \small{conv+bn+act} & 512 & (3,3) & (2,2) & \small{LeakyReLU}\\
\hline
\small{6} & \small{conv+bn+act} & 512 & (3,3) & (2,2) & \small{LeakyReLU}\\
\hline
\small{7} & \small{conv+bn+act} & 512 & (3,3) & (2,2) & \small{LeakyReLU}\\
\hline
\small{8} & \small{conv+bn+act} & 512 & (3,3) & (2,2) & \small{LeakyReLU}\\
\hline
\small{9} & \small{conv+bn+act} & 512 & (3,3) & (1,1) & \small{LeakyReLU}\\
\hline
\small{10} & \small{conv+bn+act} & 512 & (3,3) & (1,1) & \small{LeakyReLU}\\
\hline
\small{11} & \small{conv+bn+act} & 512 & (3,3) & (1,1) & \small{LeakyReLU}\\
\hline
\small{12} & \small{conv+bn+act} & 256 & (3,3) & (1,1) & \small{LeakyReLU}\\
\hline
\small{13} & \small{conv+bn+act} & 128 & (3,3) & (1,1) & \small{LeakyReLU}\\
\hline
\small{14} & \small{conv+bn+act} & 64 & (3,3) & (1,1) & \small{LeakyReLU}\\
\hline
\small{15} & \small{conv+act} & 3 & (3,3) & (1,1) & \small{LeakyReLU}\\
\hline
\end{tabular}
\end{center}
\end{table}

\subsection{Deep Polarimetric HDR Reconstruction} \label{ssec:polarimetric_hdr_reconstruction}
As discussed in Section~\ref{ssec:polarization_theory}, the polarization state of light can be quantified by DoLP, which determines the portion of light that is polarized for a given pixel. Where DoLP is low, it indicates a low measure of polarization that can be attributed to similar pixel values or pixel saturation among the four polarization images. Such regions correspond to a higher weight in the CNN, and thus are regions that the CNN will direct its learning upon. In other words, the polarimetric information provided by DoLP weighs the contribution of the features extracted at each convolutional layer, and guides the CNN to estimate HDR for all properly exposed and poorly exposed regions. On the other hand, where DoLP is high, it indicates large pixel variations among the four polarization images, where existing rich and reliable polarimetric cues are available for HDR recovery. Equipped with this domain knowledge about polarization, it allows us to use this polarimetric cue as a prior in the form of soft mask for our feature masking mechanism, to reconstruct HDR using polarization images.

\begin{figure}[t!]
\centering
\includegraphics[width=0.6\linewidth]{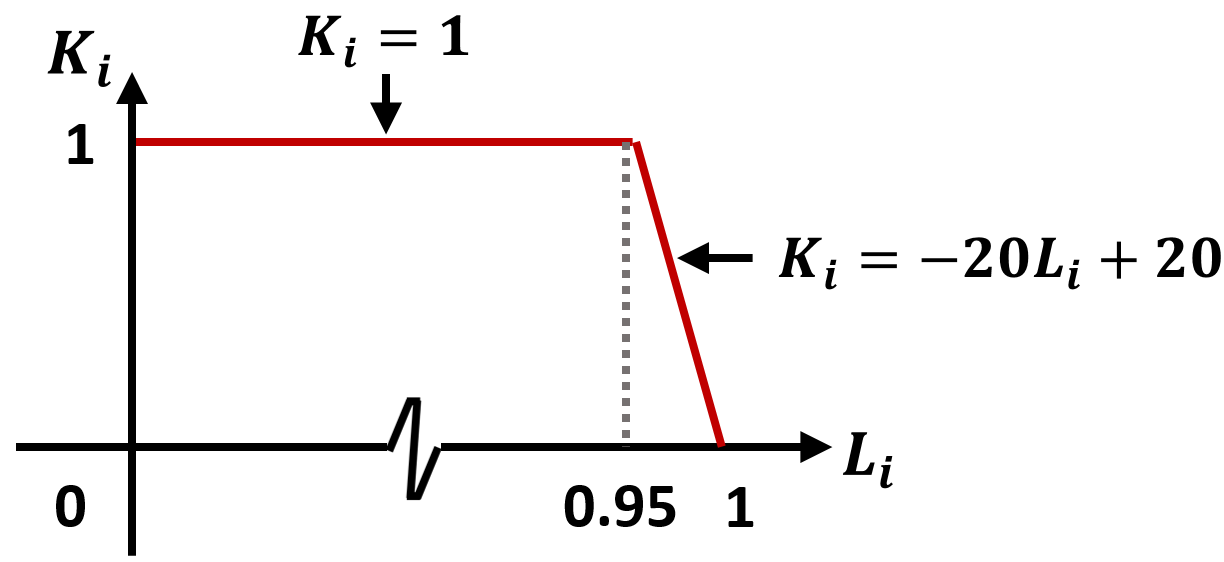}
\caption{The defined function \cite{Santos} measures how properly exposed a pixel is. The value 1 indicates the pixel as properly exposed, while 0 indicates the pixel as completely over-exposed. In our implementation, we set the threshold at 0.95.}
\label{fig:DSSHDR_K_l}
\end{figure}

In the proposed DPHR method, the first step is to stack the four LDR polarization images $L_i$, $i$ = 1,2,3,4 (512x512x12), to account for the contributions from filters oriented at 0$^{\circ}$, 45$^{\circ}$, 90$^{\circ}$, and 135$^{\circ}$, respectively for input to the CNN. The values in $L_i$ are in the range of [0,1]. Then we use $L_i$ to compute the feature mask $M_{l,i}$ where the subscripts $l$ and $i$ are the layer in the CNN and the index of each polarizer orientation image, respectively. Specifically, we compute the feature mask $M_{1,i}$ for input to the feature masking mechanism. The feature mask is constructed where if the pixel value of $L_i$ is properly exposed and DoLP is low, then the reconstruction is dominated by the predicted image $H_d$. However, if $L_i$ is poorly exposed and DoLP is high, then the reconstruction is dominated by the traditional model based method $H_t$. Then for $L_i$ and DoLP values that lie somewhere in between, the reconstruction is shared by the $H_d$ and $H_t$ images. The feature mask $M_{1,i}$ is defined as follows:
\begin{equation}
M_{1,i}=\frac{\rho + K_i}{max(\rho + K_i)}
\label{eqn:DSSHDR_mask_M}
\end{equation}

\noindent where the subscript $1$ indicates the first layer in the CNN ($l$ = 1). The values in $K_i$ are in the range of [0,1], and indicate the proper exposedness of each input pixel (of $L_i$) based on the pixel intensity, as shown in Fig.~\ref{fig:DSSHDR_K_l}. The value 0 in $K_i$ indicates the input pixel is completely over-exposed. Therefore, $M_{1,i}$ computes the well exposedness of each input pixel based on the combination of polarization and intensity information. The values in $M_{1,i}$ are in the range of [0,1]. The value 0 in $M_{1,i}$ indicates the feature is computed from low polarization and poorly exposed pixels, and thus represents invalid content that require reconstruction by the CNN. Then the feature mask is updated at each convolutional layer $l$ to obtain $M_{l,i}$. As a result, the reconstruction is achieved by using the mask $M_{l,i}$ to reduce the magnitude of the features generated from the invalid content, by updating the feature maps $X_{l,i}$ extracted at each convolutional layer as follows:
\begin{equation}
X_{l+1,i} = X_{l,i} \cdot M_{l,i}
\label{eqn:DSSHDR_Z_l}
\end{equation}

\noindent where, with the abuse of notation, the multiplication sign $\cdot$ means the pixel-wise multiplication where appropriate throughout the paper. In addition, the mask at each layer is computed by applying the convolutional filter to the mask at the previous layer. Since the masks are in the range of [0,1] and weighs the contributions of the features, the magnitude of the filters are not important. As a result, the filter weights are normalized before they are used in convolution with the mask as follows:
\begin{equation}
M_{l+1, i} = \bigg(\frac{|W_{l,i}|}{{\left\lVert W_{l,i} \right\rVert}_1 + \epsilon}\bigg) \ast M_{l,i}
\label{eqn:DSSHDR_M_l+1}
\end{equation}

\noindent where ${\left\lVert \cdot \right\rVert}_1$ is the $l_1$ function,  $|\cdot|$ is the absolute operator, $\ast$ is the convolution operator, and $\epsilon = 10^{-6}$ is a small constant added to avoid division by 0. 

We define our loss function as follows:
\begin{equation}
\mathcal{L} = \lambda_1 \mathcal{L}_r + \lambda_2 \mathcal{L}_p
\label{eqn:DSSHDR_loss_Lr_Lp}
\end{equation}

\noindent where $\mathcal{L}_r$ is an HDR reconstruction loss and $\mathcal{L}_p$ is a perceptual loss. $\lambda_1 = 6.0$ and $\lambda_2 = 1.0$ in our implementation. 

The reconstruction loss computes the pixel-wise $l_1$ distance between the ground truth image $H_g$ and predicted image $H_d$ for the invalid content defined as follows:
\begin{equation}
\mathcal{L}_r = {{\left\lVert (1 - M_1) \cdot (H_g - H_d) \right\rVert}_1}
\label{eqn:DSSHDR_loss_Lr}
\end{equation}

\noindent where $M_1$ is the mean of the four input feature masks, thereby accounting for the weights from each polarization filter.  

The perceptual loss has been demonstrated useful to improve visual quality \cite{Sun_2020_LearnedOpticHDR, Santos}, and its key idea is to enhance the similarity in feature space between the ground truth and predicted images. Our perceptual loss is a combination of the VGG loss $\mathcal{L}_v$ and style loss $\mathcal{L}_s$ as follows:
\begin{equation}
\mathcal{L}_p = \lambda_3 \mathcal{L}_v + \lambda_4 \mathcal{L}_s
\label{eqn:DSSHDR_loss_Lp}
\end{equation}

\noindent where $\lambda_3 = 1.0$ and $\lambda_4 = 120.0$ in our implementation. 

The VGG loss is defined as follows:
\begin{equation}
\mathcal{L}_v = \sum_{l} {{\left\lVert \phi_l(H_g) - \phi_l(H_d) \right\rVert}_1}
\label{eqn:DSSHDR_loss_Lv}
\end{equation}

\noindent where $\phi_l$ is the feature map from the $l$-th layer of the pretrained VGG-19. Then to recover more vivid textures, we add the style loss to (\ref{eqn:DSSHDR_loss_Lp}), and the style loss is defined as follows: 
\begin{equation}
\mathcal{L}_s = \sum_{l} {{\left\lVert G_l(H_g) - G_l(H_d) \right\rVert}_1}
\label{eqn:DSSHDR_loss_Ls}
\end{equation}

\noindent where $G_l$ is the Gram matrix \cite{Gatys_2016_CVPR} applied on the feature map at the $l$-th layer of the pretrained VGG-19. For the perceptual loss, we extract feature maps from pool1, pool2 and pool3 layers of the VGG-19 network.

\begin{figure}[t!]
\centering
\includegraphics[width=0.95\linewidth]{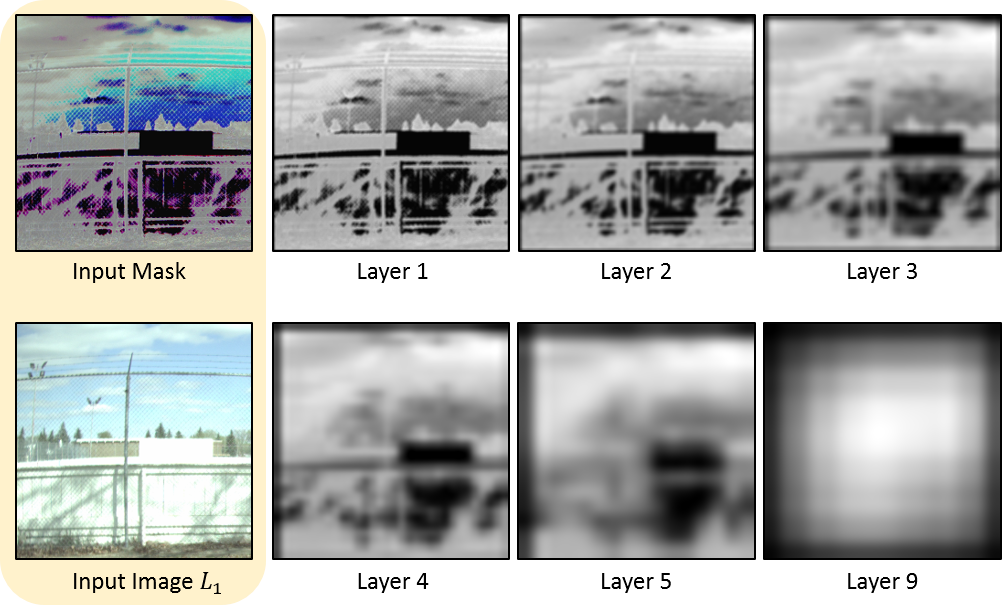}
\caption{Visualization of the masks $M_{l,i=1}$ at different layers of the network, where the dark pixels identify invalid regions with low polarization and poorly exposed pixels that require CNN to direct its learning. We can observe that the masks become blurrier and more uniform in the deeper layers, as the receptive fields of the convolution filters become larger.}
\label{fig:chapt5_mask_M_l}
\end{figure}

Fig.~\ref{fig:chapt5_mask_M_l} shows the mask $M_{l,i=1}$ at different layers of the network. We can observe from the input image that the building and the fence regions are over-exposed. We can also observe from the input mask that these regions have a low measure of polarization. The input mask informs the network where the invalid features are for the CNN to direct its learning upon. Therefore, the building and the fence regions consistently indicate a lower polarization and a poorer exposed pixel (i.e., darker pixel) compared to other regions in the masks at each convolutional layer. Also observed is that the masks become blurrier and more uniform in the deeper layers, as the receptive fields of the convolution filters become larger.

\subsection{HDR Formulation} \label{ssec:hdr_formulation}
With the CNN predicted image $H_d$, we formulate the final HDR image $H$ through HDR integration as follows:
\begin{equation}
H = \alpha \cdot H_t + (1 - \alpha) \cdot H_d
\label{eqn:DSSHDR_inference}
\end{equation}

\noindent where $\alpha$ is as follows:
\begin{equation}
\alpha = \frac{\rho}{\rho + (1 - M_1)}
\label{eqn:DSSHDR_inference_alphagamma}
\end{equation}

\noindent In (\ref{eqn:DSSHDR_inference}), $H_t$ is computed as follows \cite{xuesong}:
\begin{equation}
H_t =\frac{\sum_{i=1}^{2}W\Big(L_{i}+L_{i+2}\Big)\Big(g\big(L_{i}\big)+g\big(L_{i+2}\big)\Big)}{\sum_{i=1}^{2}W\Big(L_{i}+L_{i+2}\Big)t_0} 
\label{eqn:I_deb}
\end{equation}

\noindent where $W$ is the Gaussian weighted function ($\sigma$ = 0.2 in our study), and $g$ is the inverse camera response function.

The HDR formulation by (\ref{eqn:DSSHDR_inference}) uses a combination of $H_t$ which is a traditional model based method, and $H_d$ which is a deep-learning-based method, to estimate HDR in all areas. $H_t$ can estimate HDR well for all areas with high DoLP \cite{xuesong}. Then we let $H_d$ to estimate HDR for all areas with low DoLP, using the predicted pixels from the CNN. Additionally, normalization is performed where necessary.

\section{Experimental Results and Discussion} \label{sec:Experiments}
In this section, we first present the experimental setups, and then evaluate the experimental results of the proposed DPHR method on our collected outdoor polarization images. Specifically, we evaluate the effectiveness of the different components in our HDR formulation. To show the capability of the proposed DPHR method, we compare the reconstructed results with state-of-the-art methods.

\begin{figure}[t!]
\centering
\includegraphics[width=0.95\linewidth]{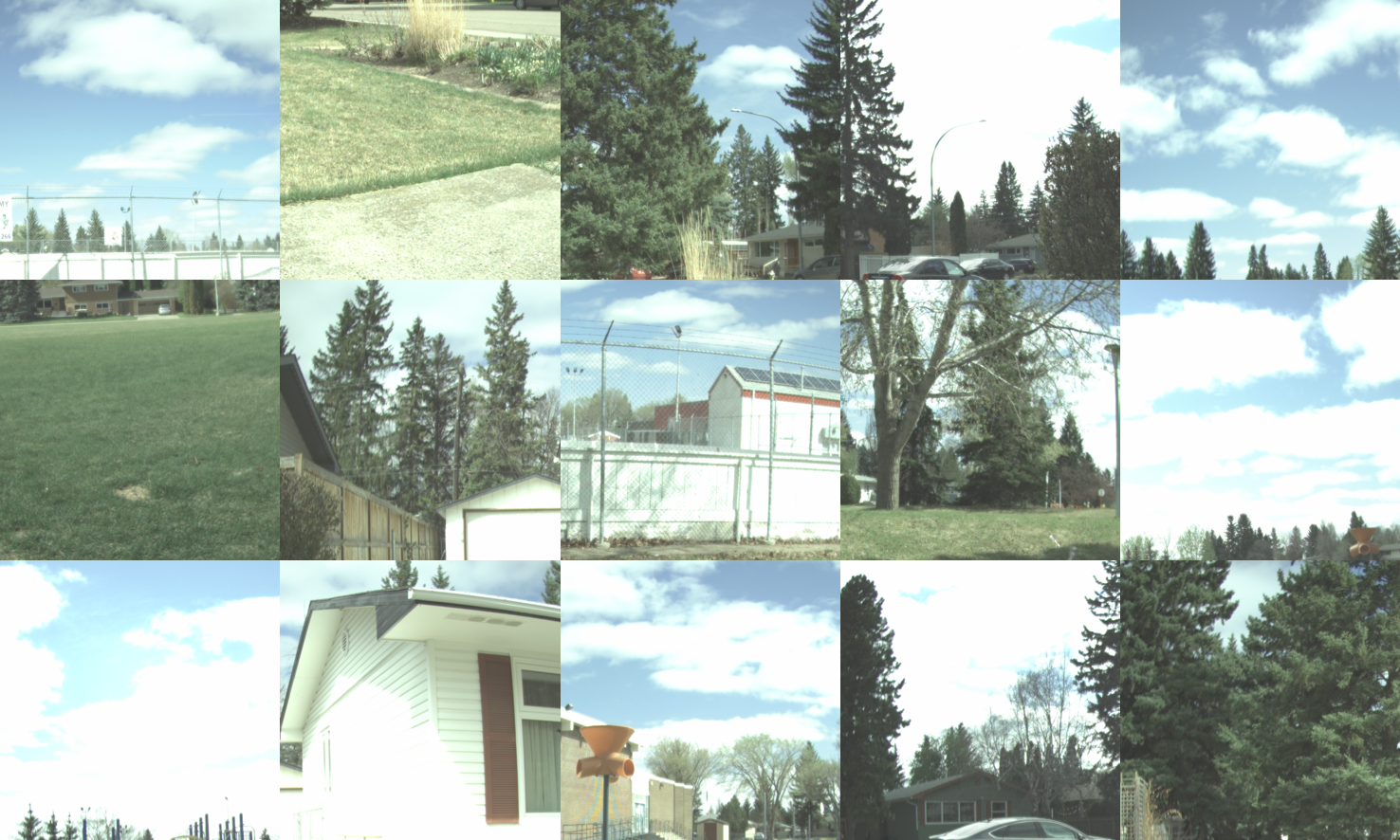}
\caption{Examples of input LDR images $L_1$ at $t_0$ = 0.769 ms from the EdPolCommunity dataset.}
\label{fig:dataset_I_cmean}
\end{figure}

\subsection{Experiment Setups}
\noindent \textbf{Dataset.} Since there is a lack of polarization images for training our network for HDR reconstruction, we have built our own dataset. We collected an outdoor dataset, under sunlight conditions, named EdPolCommunity. The dataset is collected by the IMX250MYR polarization camera \cite{Polcam}. For each scene, four high-resolution 1024x1224 colored polarization images is captured at 17 exposure times ($t_0$ = 0.03, 0.045, 0.068, 0.101, 0.152, 0.228, 0.342, 0.513, 0.769, 1.153, 1.73, 2.595, 3.592, 5.839, 8.758, 13.137, 19.705 ms). Then four 512x512 patches are cropped from each image for improved readability. In total, the EdPolCommunity dataset has 736 pairs of ground truth HDR and LDR images available to train and test the network. We follow a 70/20/10 practice to split our dataset into train/val/test sets. The images used for training are not used in the evaluation. Fig.~\ref{fig:dataset_I_cmean} shows examples of input LDR images $L_1$ at $t_0$ = 0.769 ms from the EdPolCommunity dataset.

To gain further insights on the EdPolCommunityOutdoor dataset, a mixture model and the expectation maximization algorithm \cite{EM_algo}  is used to fit the DoLP distribution. The mixture model is given by:
\begin{equation}
f = w \cdot \gamma + (1-w) \cdot U
\label{eqn:DoP_EdPolCommunityOutdoor_mixture}
\end{equation}

\noindent where $w$ is the weight of the distributions, $\gamma$ is the Gamma distribution, and $U$ is the uniform distribution. For the EdPolCommunityOutdoor dataset: $w$ = 0.934, $\gamma_{\alpha}$ = 6.264, $\gamma_{\beta}$ = 0.023, $U_{start}$ = 0.0 and $U_{end}$ = 1.0. We note that (\ref{eqn:DoP_EdPolCommunityOutdoor_mixture}) is a general case that considers the presence of shot noise which is prevalent in poorly-lit scenes. For the EdPolCommunityOutdoor dataset, $f$ in (\ref{eqn:DoP_EdPolCommunityOutdoor_mixture}) is predominantly weighted by the Gamma distribution since its weight $w$ is 0.934, and thus the data can be explained well with only the Gamma distribution. The negative likelihood calculated for the mixture model is -548.895. Additionally, the acquisition settings are summarized in Table~\ref{tab:chapt3_capture_settings}.

\begin{table}[t]
\caption{Polarized capture settings}
\vspace{-1.0\baselineskip}
\begin{center}
\label{tab:chapt3_capture_settings}
\begin{tabular}{c|c|c|c}
\hline
\small{\textbf{Aperture}} & \small{16} & \small{\textbf{FOV}} & \small{58$^{\circ}$x49$^{\circ}$x73$^{\circ}$}\\ 
\small{\textbf{Focal length}} & \small{8} & \small{\textbf{Pixel format}} & \small{Polarized 8}\\ 
\small{\textbf{Gain}} & \small{0} & \small{\textbf{Pixel range}} & \small{0-255}\\ 
\small{\textbf{Black level}} & \small{2.22} & \small{\textbf{ADC bit depth}} & \small{12}\\
\small{\textbf{Sensor size}} & \small{2464x2056} & \small{\textbf{Frame rate}} & \small{73}\\
\small{\textbf{Image size}} & \small{2448x2048} & \small{\textbf{\# Exposures}} & \small{17}\\
\hline
\end{tabular}
\end{center}
\end{table}

\smallskip
\noindent \textbf{Implementation Details.} The network adopts a U-Net architecture \cite{ronneberger2015unet}. All the encoder layers use the leaky ReLU activation ($\alpha$ = 0.2), while we use ReLU in all the decoder layers, except for the last one, which has a linear activation. Also the feature masking strategy is used in all the convolutional layers. The framework is implemented using PyTorch and trained on NVIDIA GeForce GTX 1080 Ti. We initialized the network using the Xavier approach. Then trained using an Adam optimizer \cite{Adam} with the default parameters $\beta_1$ = 0.9 and $\beta_2$ = 0.999, and mini-batch size of 4.

\smallskip
\noindent \textbf{Evaluation Metrics.} To evaluate the quality of the HDR images, we used the popular HDR-VDP2 \cite{hdrvdp2} metric, and adopted the means squared error (MSE) metric as used in other works \cite{Liu, Santos}. We normalize the predicted and reference ground truth HDR images \cite{Marnerides}. Then to account for the human visual system's sensitivity to lighting at different luminance values, we apply perceptual uniform (PU) encoding to calculate PU-PSNR and PU-SSIM metrics \cite{Mantiuk2021PU21}, to further evaluate the accuracy of the HDR images.

\begin{figure*}[t!]
\centering
\begin{subfigure}{.18\textwidth}
  \centering
  \includegraphics[width=0.95\linewidth]{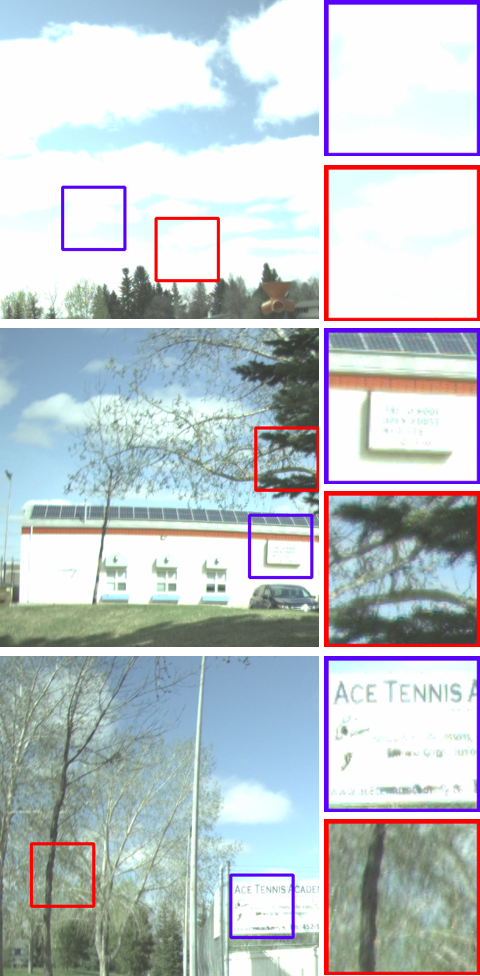}
  \caption{$L_1$}
  \label{fig:DSSHDR_ablation_model_in_pol0}
\end{subfigure}%
\begin{subfigure}{.18\textwidth}
  \centering
  \includegraphics[width=0.95\linewidth]{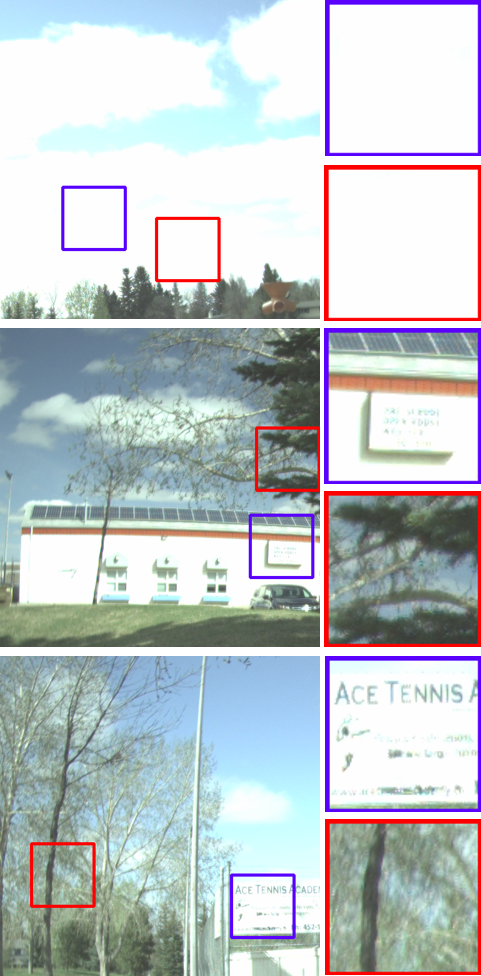}
  \caption{$L_2$}
  \label{fig:DSSHDR_ablation_model_in_pol45}
\end{subfigure}%
\begin{subfigure}{.18\textwidth}
  \centering
  \includegraphics[width=0.95\linewidth]{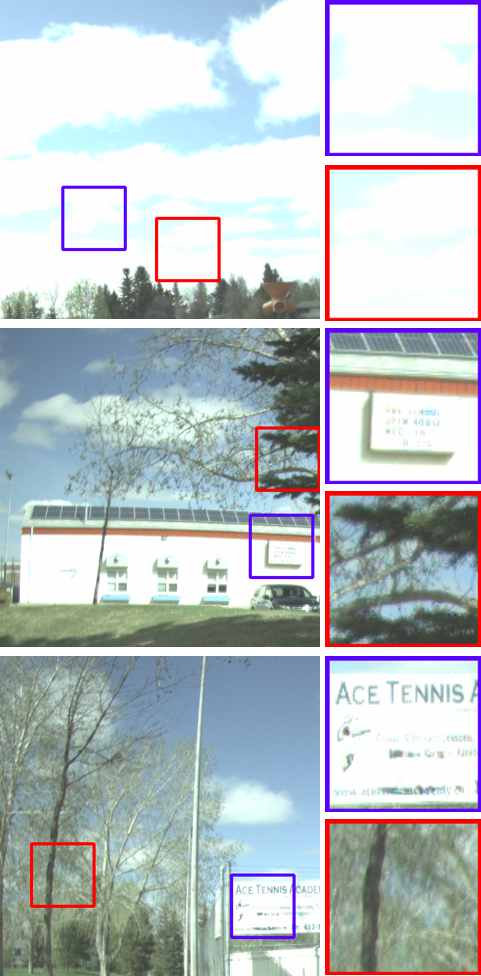}
  \caption{$L_3$}
  \label{fig:DSSHDR_ablation_model_in_pol90}
\end{subfigure}%
\begin{subfigure}{.18\textwidth}
  \centering
  \includegraphics[width=0.95\linewidth]{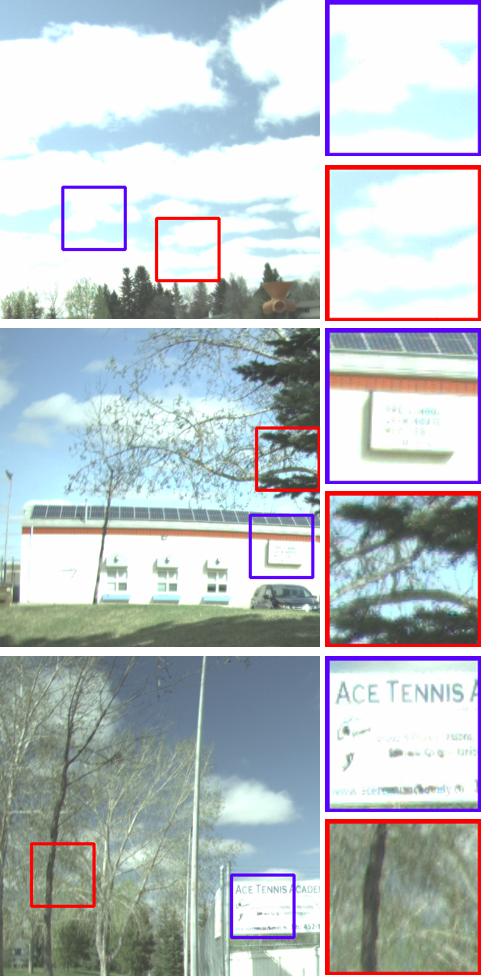}
  \caption{$L_4$}
  \label{fig:DSSHDR_ablation_model_in_pol135}
\end{subfigure}%
\centering
\vskip 0pt
\begin{subfigure}{.14\textwidth}
  \centering
  \includegraphics[width=0.95\linewidth]{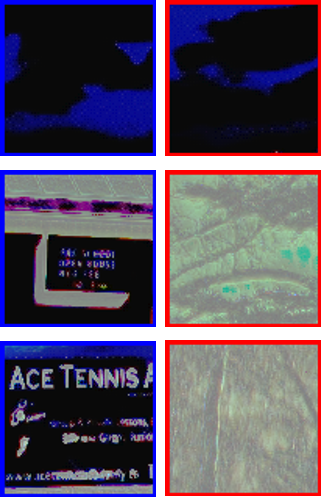}
  \caption{$M_{1,1}$}
  \label{fig:DSSHDR_ablation_model_inputmaskN_pol0}
\end{subfigure}%
\begin{subfigure}{.14\textwidth}
  \centering
  \includegraphics[width=0.95\linewidth]{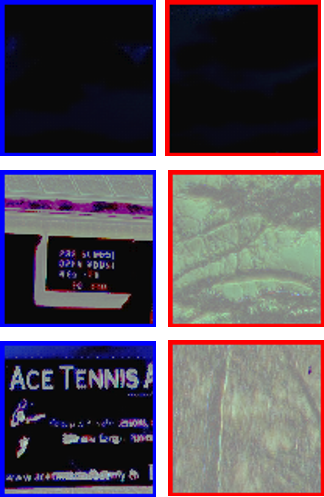}
  \caption{$M_{1,2}$}
  \label{fig:DSSHDR_ablation_model_inputmaskN_pol45}
\end{subfigure}%
\begin{subfigure}{.14\textwidth}
  \centering
  \includegraphics[width=0.95\linewidth]{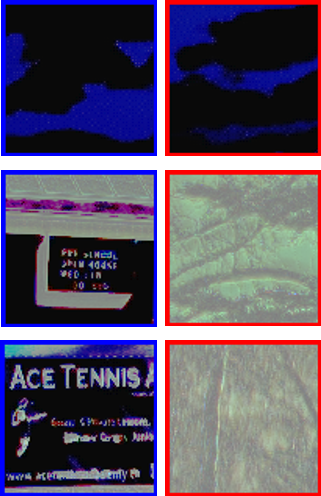}
  \caption{$M_{1,3}$}
  \label{fig:DSSHDR_ablation_model_inputmaskN_pol90}
\end{subfigure}%
\begin{subfigure}{.14\textwidth}
  \centering
  \includegraphics[width=0.95\linewidth]{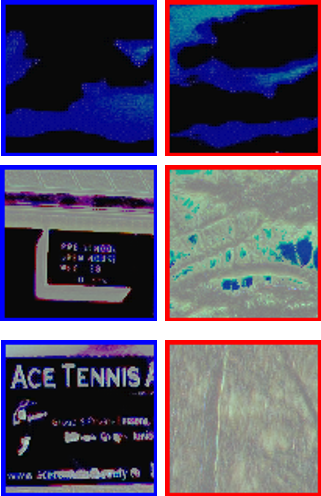}
  \caption{$M_{1,4}$}
  \label{fig:DSSHDR_ablation_model_inputmaskN_pol135}
\end{subfigure}%
\centering
\vskip 0pt
\begin{subfigure}{.18\textwidth}
  \centering
  \includegraphics[width=0.95\linewidth]{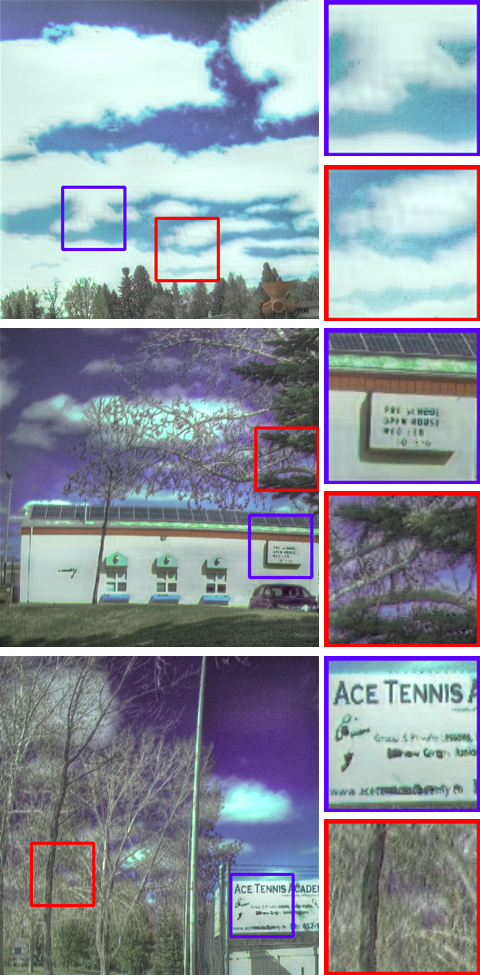}
  \caption{DPHR w/o $H_t$}
  \label{fig:chapt5_DSSHDR_ablation_model_t3}
\end{subfigure}%
\begin{subfigure}{.18\textwidth}
  \centering
  \includegraphics[width=0.95\linewidth]{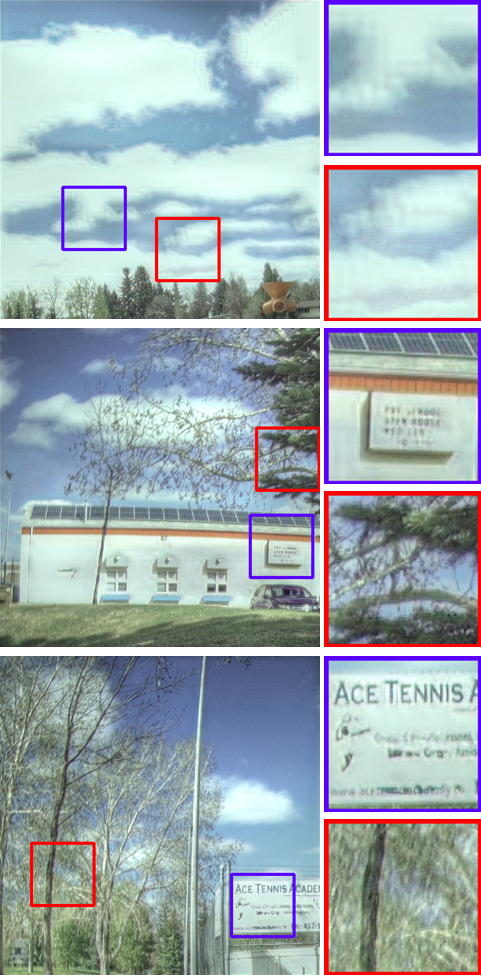}
  \caption{DPHR}
  \label{fig:chapt5_DSSHDR_ablation_model_t2nt3n}
\end{subfigure}%
\begin{subfigure}{.18\textwidth}
  \centering
  \includegraphics[width=0.95\linewidth]{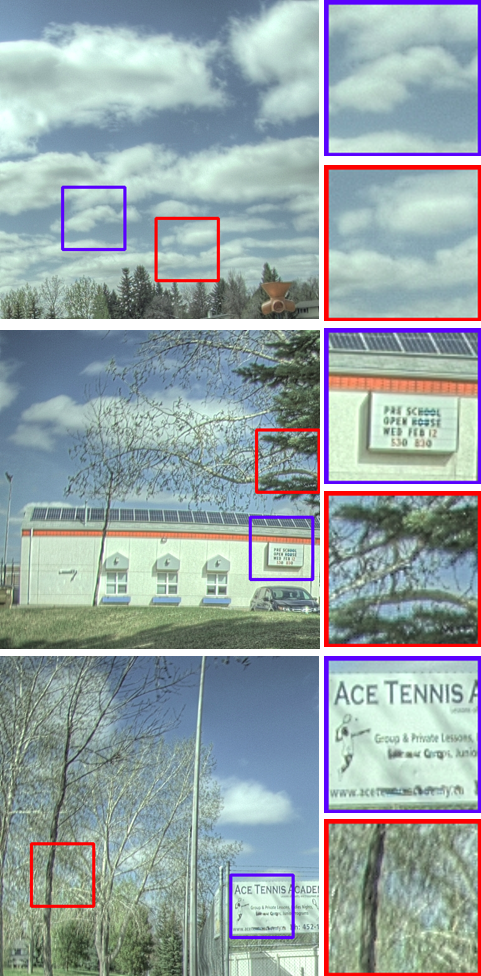}
  \caption{ground truth}
  \label{fig:chapt5_DSSHDR_ablation_model_gt}
\end{subfigure}%
\caption{Qualitative results on variants of DPHR. The DPHR model formulates HDR using a combination of the traditional model based method $H_t$ and the deep-learning-based method $H_d$, to estimate HDR in all areas. In contrast to $L_i$ and DPHR w/o $H_t$, DPHR obtains results with better details that are more closely matched to the ground truth. The input mask $M_{1,i}$ identifies dark pixels as invalid regions that require CNN to direct its learning. The Photomatix Enhanced tone-mapping operator is used.}
\label{fig:chapt5_DSSHDR_ablation_model}
\end{figure*}

\subsection{Evaluation of the HDR Formulation}
In this experiment, we validate the model design choice for the HDR formulation, which is performed by comparing the following variants of the HDR formulation: 
\begin{enumerate}[label=(\alph*)]
\item \textbf{DPHR} (\textit{i.e., $H = \alpha \cdot H_t + (1 - \alpha) \cdot H_d$}). The full HDR formulation, which is the combination of traditional model based and deep-learning-based HDR reconstruction results.

\item \textbf{DPHR w/o} \bm{$H_t$} (\textit{i.e., $H = H_d$}). We remove the traditional model based result in this variant, and obtain the final HDR image $H$ using our deep-learning-based result to show the effect of polarimetric information in comparison with the regular images in traditional deep-learning methods.
\end{enumerate}

The HDR formulation with (a) is effective for estimating HDR values. As shown in Fig.~\ref{fig:chapt5_DSSHDR_ablation_model}, compared with the input $L_i$, (b) can restore more details as our CNN uses the polarimetric information as a prior. In particular, the CNN itself can outperform the traditional deep-learning approach by Santos et al. \cite{Santos}, where the (b) variant provides better quantitative results than \cite{Santos}, as shown in Table~\ref{tab:chapt5_DSSHDR_soa} (seventh row). However, (b) suffers color distortions, and is an overall darker image with limited recovery of details in the poorly exposed regions. On the other hand, (a) can recover better details in both under- and over-exposed regions, and alleviate color distortions and visible artifacts. An explanation for the removal of color distortions in (a) is that DoLP is a mask shared among the colored channels, and thus can enforce consistency between images which helps to remove the colorization artifacts for a better HDR reconstruction. The quantitative results are shown in Table~\ref{tab:chapt5_DSSHDR_ablation_model}, and is aligned with the observation that (a) achieves a better performance. The key difference between these two variants is the effect of DoLP. In the (a) variant, DoLP is used as a strong accurate prior to recover HDR, where for regions with high DoLP the traditional model based method recovers HDR details well compared to the deep-learning-based method. Therefore, we achieve HDR formulation using the (a) variant.

In addition, the results from both (a) and (b) variants can outperform a recent deep-learning-based HDR reconstruction approach developed using the polarization camera \cite{ting2021}. Unlike \cite{ting2021} that uses only the polarization images to achieve HDR reconstruction, and thus neglects the polarimetric cue available from the polarization camera, the (b) variant integrates the polarimetric cue as a prior in the form of soft mask for our feature masking mechanism, to reconstruct better HDR images than \cite{ting2021} as shown in Table~\ref{tab:chapt5_DSSHDR_soa} (fifth row). Then the (a) variant further enhances the reconstruction results of \cite{ting2021} in Table~\ref{tab:chapt5_DSSHDR_soa} by integrating a prior that the traditional model based method $H_t$ is able to recover well for regions with high DoLP to generate the final HDR image.

\begin{table}[t!]
\caption{Quantitative results on variants of DPHR (the higher the better, except for MSE)}
\begin{center}
\label{tab:chapt5_DSSHDR_ablation_model}
\setlength\tabcolsep{1.1pt} 
\begin{tabular}{c|c|c|c|c}
\hline
& \small{PU-PSNR} & \small{PU-SSIM} & \small{HDR-VDP2} &\small{MSE}\\ 
\hline
\footnotesize{DPHR w/o $H_t$} & 20.34$\pm$1.83 & 0.83$\pm$0.044 & 51.13$\pm$2.94 & 0.0357$\pm$0.0029\\
\hline
\footnotesize{\textbf{DPHR}}
& \textbf{30.59$\pm$3.19} & \textbf{0.94$\pm$0.068} & \textbf{56.15$\pm$5.28} &\textbf{0.0081$\pm$0.0006}\\
\hline
\end{tabular}
\end{center}
\end{table}

\begin{figure}[t!]
\centering
\begin{subfigure}{.12\textwidth}
  \centering
  \includegraphics[width=0.95\linewidth]{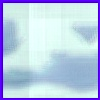}
  \caption{ $\mathcal{L}_r$}
  \label{fig: DSSHDR_ablation_loss_Lronly }
\end{subfigure}%
\begin{subfigure}{.12\textwidth}
  \centering
  \includegraphics[width=0.95\linewidth]{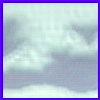}
  \caption{ $\mathcal{L}_p$}
  \label{fig: DSSHDR_ablation_loss_Lponly}
\end{subfigure}%
\begin{subfigure}{.12\textwidth}
  \centering
  \includegraphics[width=0.95\linewidth]{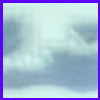}
  \caption{ $\mathcal{L}_r + \mathcal{L}_p$}
  \label{fig: DSSHDR_ablation_loss_LrLp }
\end{subfigure}%
\begin{subfigure}{.12\textwidth}
  \centering
  \includegraphics[width=0.95\linewidth]{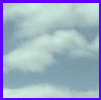}
  \caption{ground truth}
  \label{fig: DSSHDR_ablation_loss_gt }
\end{subfigure}%
\caption{Qualitative results on different loss functions, where our choice of $\mathcal{L}_r + \mathcal{L}_p$ loss presents color consistency and preserves details that are better matched with the ground truth.The Photomatix Enhanced tone-mapping operator is used.}
\label{fig:chapt5_DSSHDR_ablation_loss}
\end{figure}

\begin{table}[t!]
\caption{Quantitative results on variants of loss function (the higher the better, except for MSE)}
\begin{center}
\label{tab:chapt5_DSSHDR_ablation_loss}
\setlength\tabcolsep{1.1pt} 
\begin{tabular}{c|c|c|c|c}
\hline
& \small{PU-PSNR} & \small{PU-SSIM} & \small{HDR-VDP2} &\small{MSE}\\ 
\hline
\footnotesize{$\mathcal{L}_r$} & 29.89$\pm$2.89 & 0.90$\pm$0.034 & 54.15$\pm$3.96 & 0.0080$\pm$0.0081\\
\footnotesize{$\mathcal{L}_p$} & 29.94$\pm$2.87 & 0.93$\pm$0.023 & 55.35$\pm$4.80 & 0.0088$\pm$0.0061\\
\hline
\footnotesize{\textbf{$\mathcal{L}_r$ + $\mathcal{L}_p$}}
& \textbf{30.59$\pm$3.19} & \textbf{0.94$\pm$0.068} & \textbf{56.15$\pm$5.28} &\textbf{0.0081$\pm$0.0006}\\
\hline
\end{tabular}
\end{center}
\end{table}

\begin{figure*}[t!]
\begin{subfigure}{.18\textwidth}
  \centering
  \includegraphics[width=0.95\linewidth]{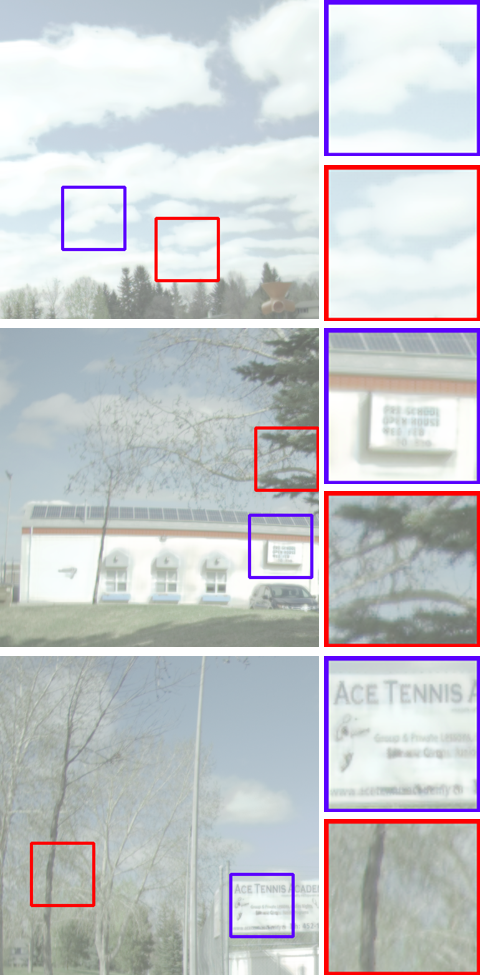}
  \caption{HDR-GAN}
  \label{fig:DSSHDR_soa_HDR-GAN}
\end{subfigure}%
\begin{subfigure}{.18\textwidth}
  \centering
  \includegraphics[width=0.95\linewidth]{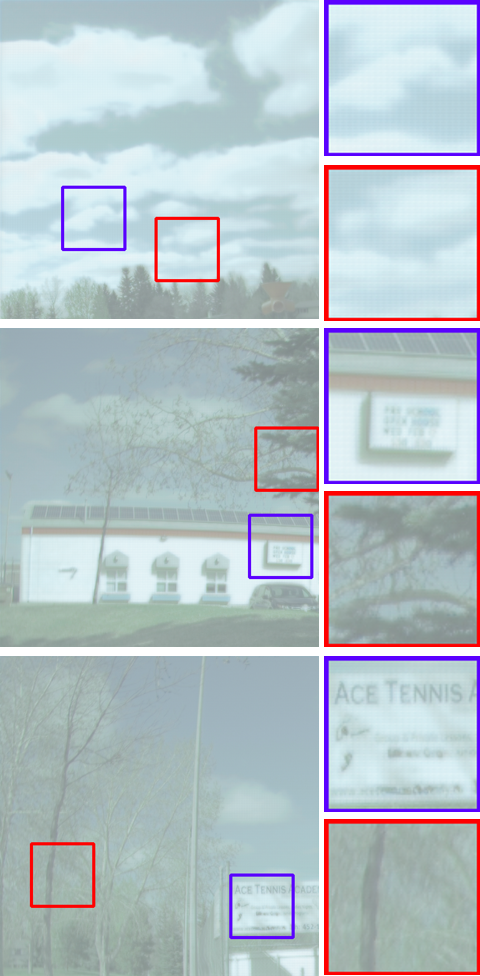}
  \caption{DeepHDR}
  \label{fig:DSSHDR_soa_DeepHDR-eccv18}
\end{subfigure}%
\begin{subfigure}{.18\textwidth}
  \centering
  \includegraphics[width=0.95\linewidth]{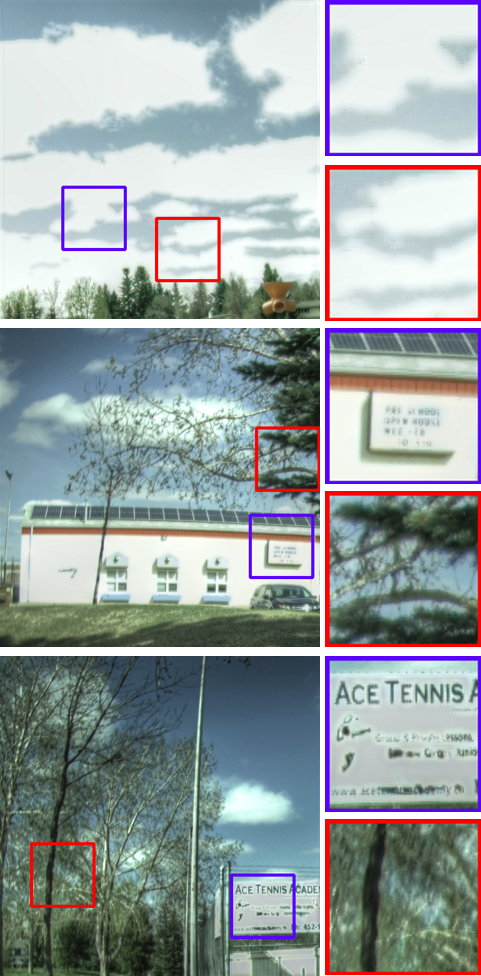}
  \caption{ENet}
  \label{fig:chapt5_DSSHDR_soa_enet}
\end{subfigure}%
\begin{subfigure}{.18\textwidth}
  \centering
  \includegraphics[width=0.95\linewidth]{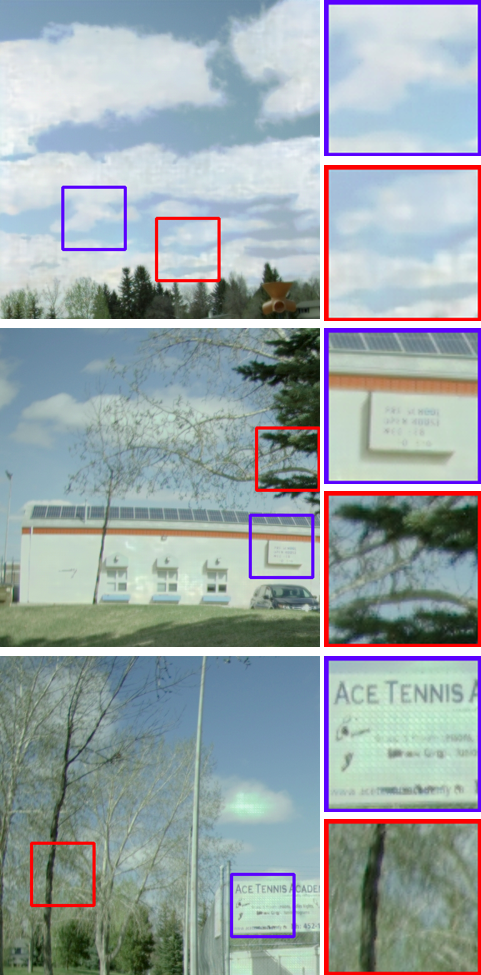}
  \caption{DrTMO}
  \label{fig:chapt5_DSSHDR_soa_drtmo}
\end{subfigure}%
\centering
\vskip 0pt
\begin{subfigure}{.18\textwidth}
  \centering
  \includegraphics[width=0.95\linewidth]{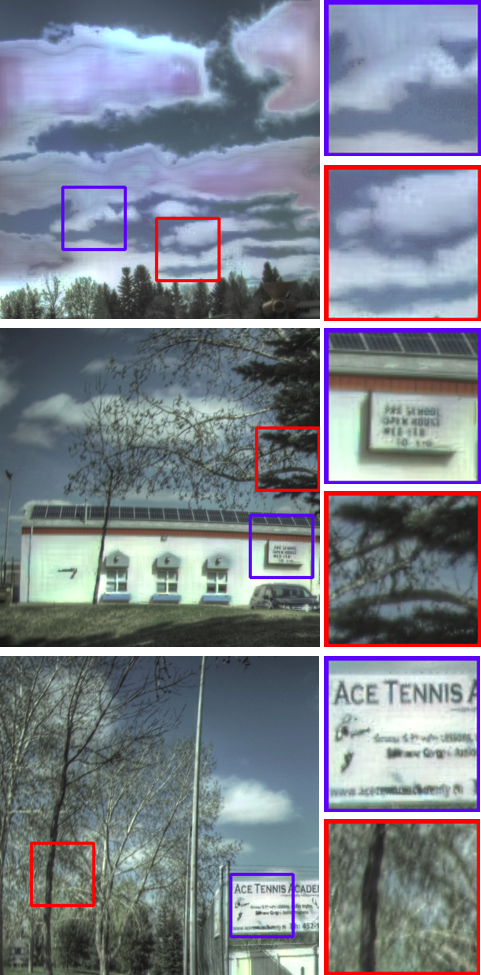}
  \caption{DRCP}
  \label{fig:chapt5_DSSHDR_soa_drcp}
\end{subfigure}%
\begin{subfigure}{.18\textwidth}
   \centering
  \includegraphics[width=0.95\linewidth]{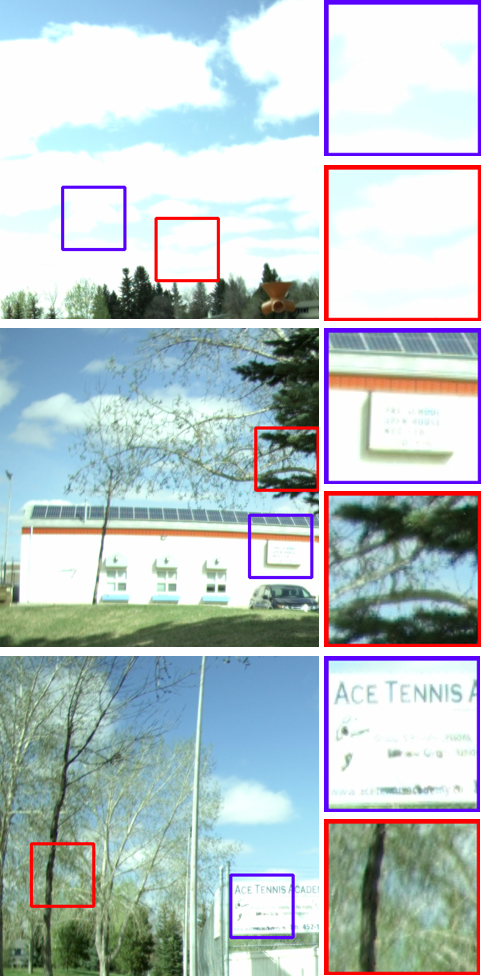}
  \caption{PHDR}
  \label{fig:chapt5_DSSHDR_soa_phdr}
\end{subfigure}%
\begin{subfigure}{.18\textwidth}
   \centering
  \includegraphics[width=0.95\linewidth]{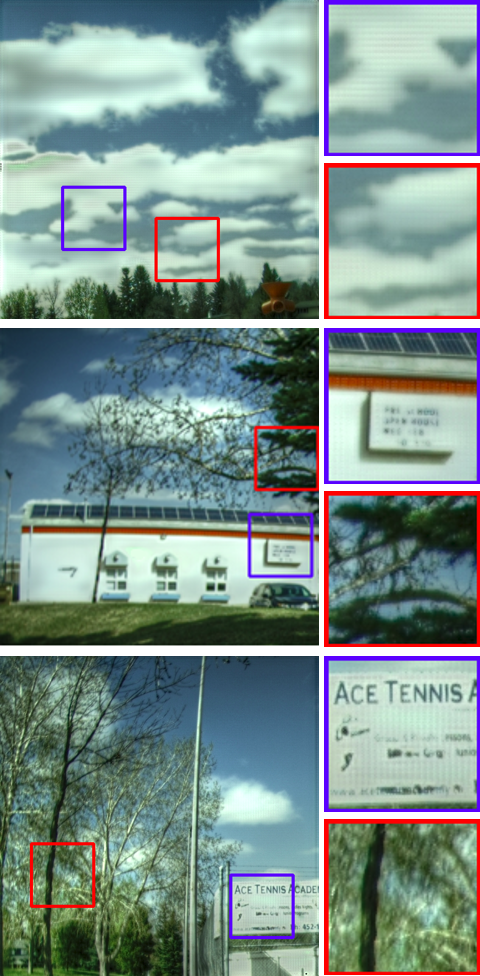}
  \caption{DPSHDR}
  \label{fig:DPSHDR}
\end{subfigure}%
\begin{subfigure}{.18\textwidth}
  \centering
  \includegraphics[width=0.95\linewidth]{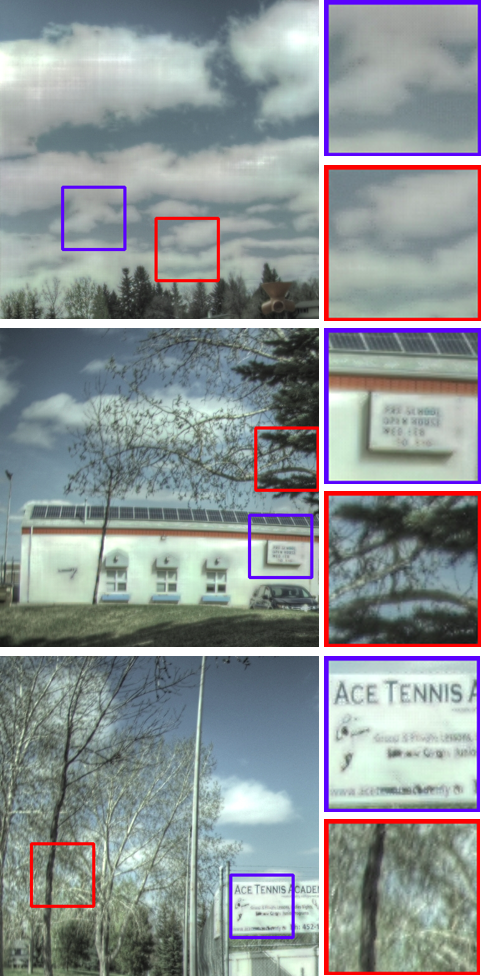}
  \caption{HDRCNN}
  \label{fig:chapt5_DSSHDR_soa_hdrcnn}
\end{subfigure}%
\centering
\vskip 0pt
\begin{subfigure}{.18\textwidth}
  \centering
  \includegraphics[width=0.95\linewidth]{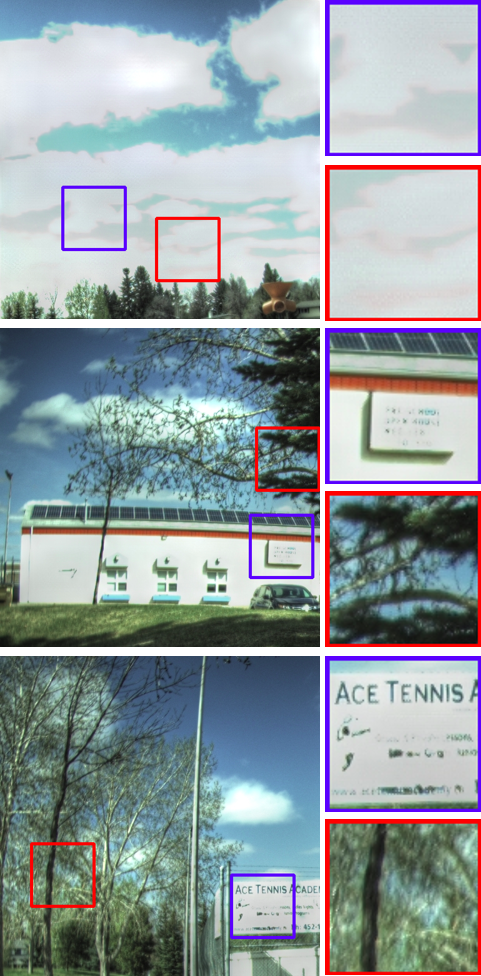}
  \caption{HDRCNN-Mask}
  \label{fig:chapt5_DSSHDR_soa_hdrcnnmask_pretrain}
\end{subfigure}%
\begin{subfigure}{.18\textwidth}
  \centering
  \includegraphics[width=0.95\linewidth]{Figures/DSSHDR_ablation_model_t2nt3n.png}
  \caption{DPHR}
  \label{fig:chapt5_DSSHDR_soa_ours}
\end{subfigure}%
\begin{subfigure}{.18\textwidth}
  \centering
  \includegraphics[width=0.95\linewidth]{Figures/DSSHDR_ablation_model_gt.png}
  \caption{ground truth}
  \label{fig:chapt5_DSSHDR_ablation_model_gt}
\end{subfigure}%
\caption{Qualitative comparative results with state-of-the-art methods. The proposed DPHR method presents color consistency, free of visible artifacts, and overall able to recover richer textures in all poorly exposed regions. This is because DPHR is designed to handle polarization images, as it integrated the polarimetric cues into the framework. The Photomatix Enhanced TMO is used.}
\label{fig:chapt5_DSSHDR_soa}
\end{figure*}


\begin{figure*}[t!]
\begin{subfigure}{.18\textwidth}
  \centering
  \includegraphics[width=0.95\linewidth]{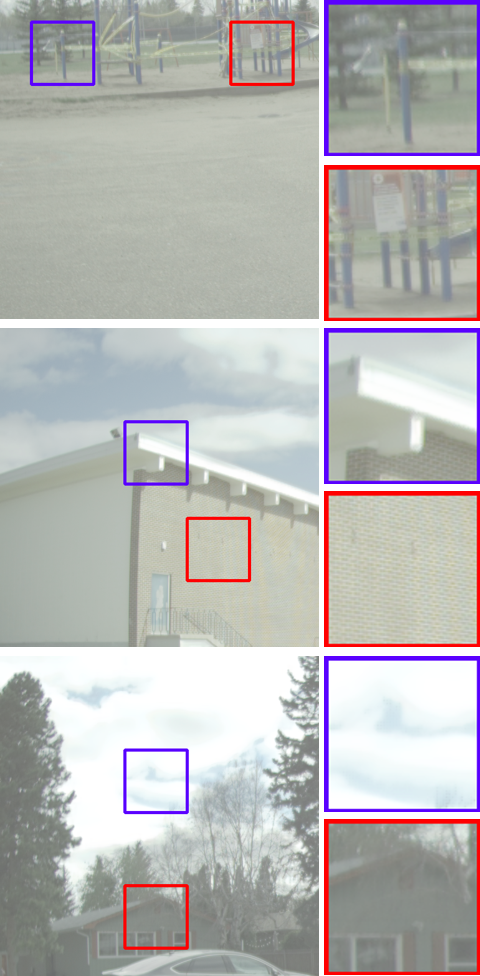}
  \caption{HDR-GAN}
  \label{fig:supp_acmtog_HDR-GAN2}
\end{subfigure}%
\begin{subfigure}{.18\textwidth}
  \centering
  \includegraphics[width=0.95\linewidth]{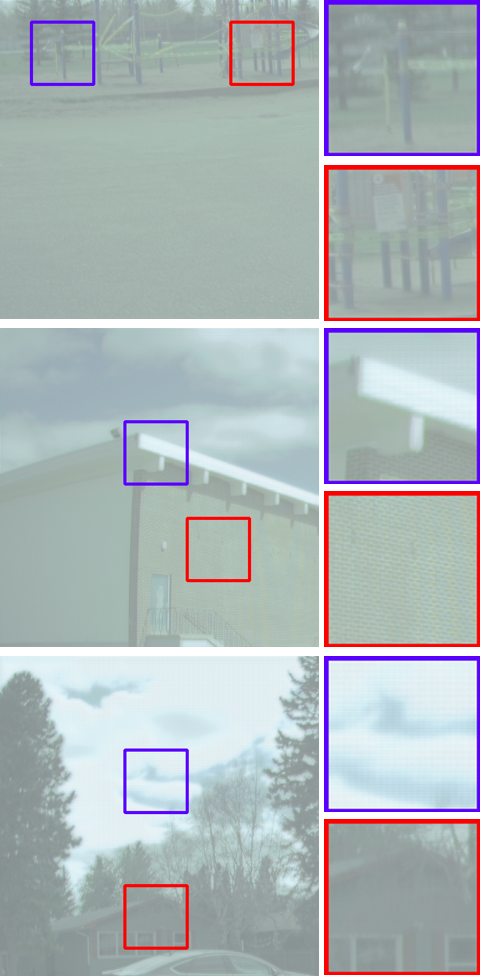}
  \caption{DeepHDR}
  \label{fig:supp_acmtog_DeepHDR-eccv18_2}
\end{subfigure}%
\begin{subfigure}{.18\textwidth}
  \centering
  \includegraphics[width=0.95\linewidth]{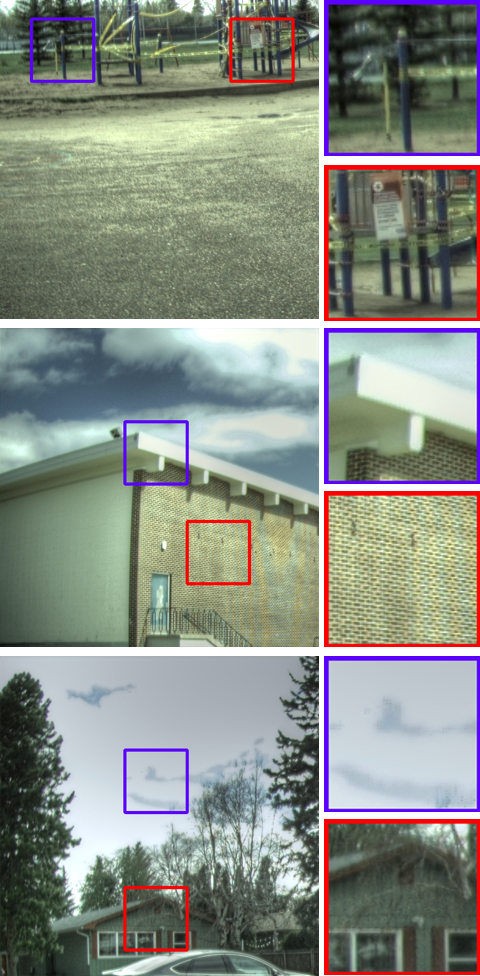}
  \caption{ENet}
  \label{fig:supp_acmtog_enet2}
\end{subfigure}%
\begin{subfigure}{.18\textwidth}
  \centering
  \includegraphics[width=0.95\linewidth]{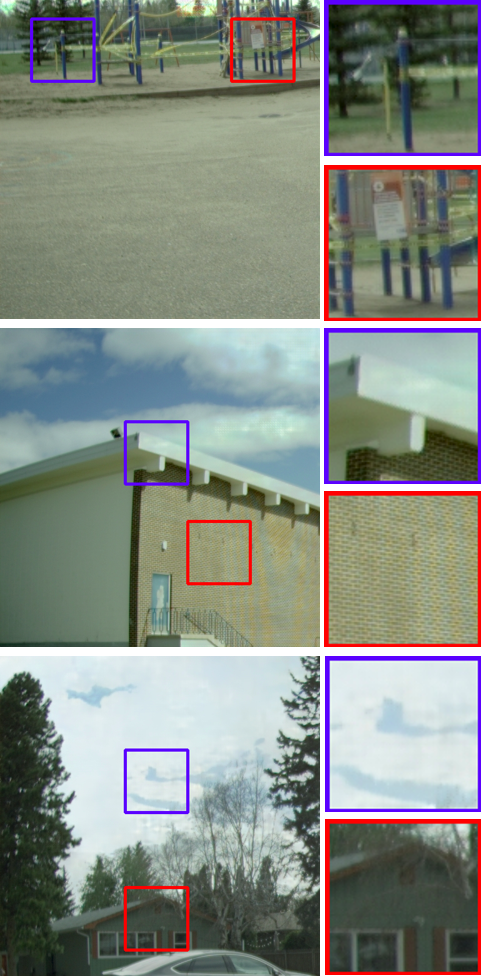}
  \caption{DrTMO}
  \label{fig:supp_acmtog_drtmo2}
\end{subfigure}%
\centering
\vskip 0pt
\begin{subfigure}{.18\textwidth}
  \centering
  \includegraphics[width=0.95\linewidth]{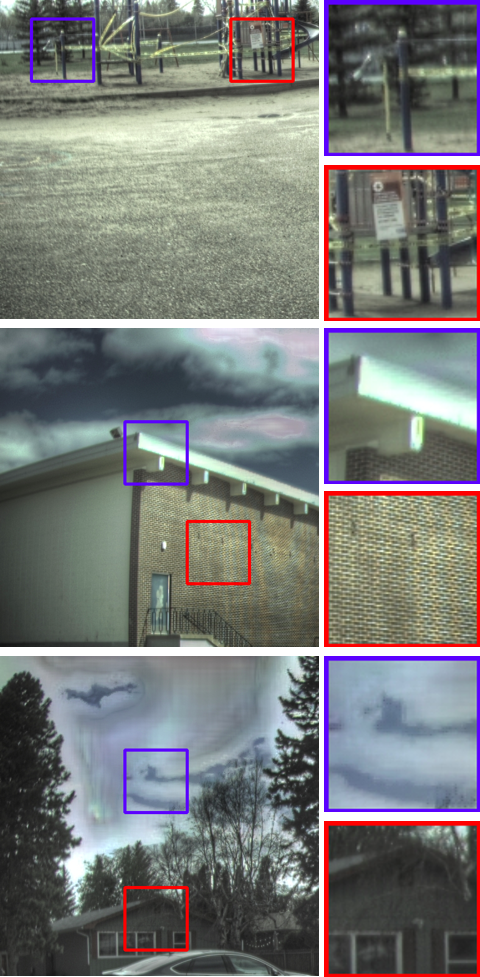}
  \caption{DRCP}
  \label{fig:supp_acmtog_drcp2}
\end{subfigure}%
\begin{subfigure}{.18\textwidth}
  \centering
  \includegraphics[width=0.95\linewidth]{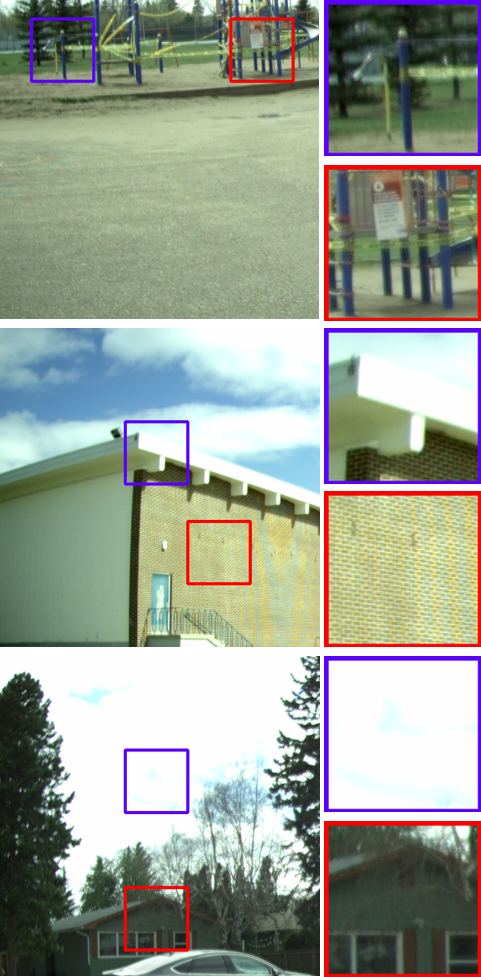}
  \caption{PHDR}
  \label{fig:supp_acmtog_phdr2}
\end{subfigure}%
\begin{subfigure}{.18\textwidth}
  \centering
  \includegraphics[width=0.95\linewidth]{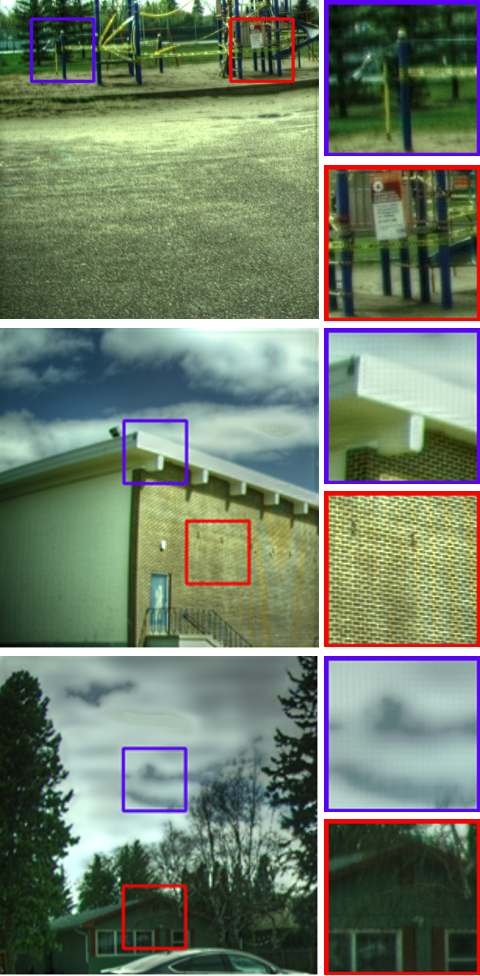}
  \caption{DPSHDR}
  \label{fig:supp_acmtog_dpshdr2}
\end{subfigure}%
\begin{subfigure}{.18\textwidth}
  \centering
  \includegraphics[width=0.95\linewidth]{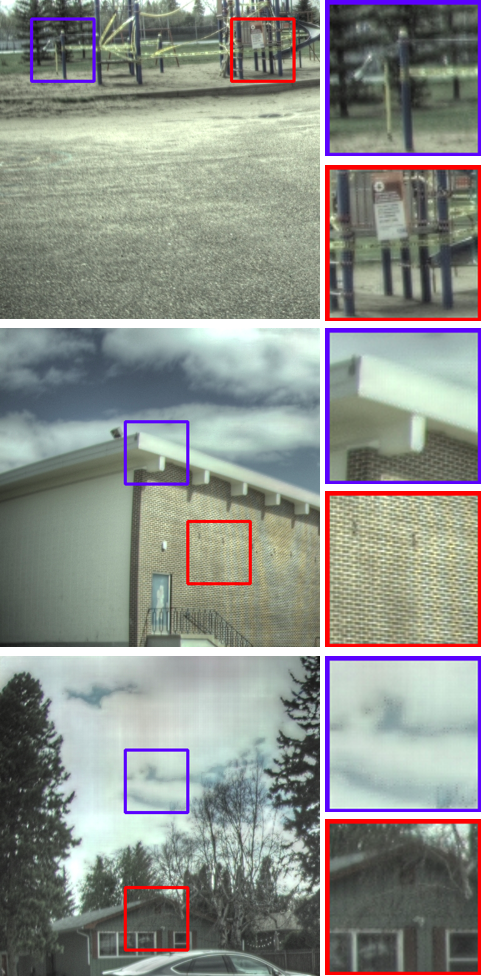}
  \caption{HDRCNN}
  \label{fig:supp_acmtog_hdrcnn2}
\end{subfigure}%
\centering
\vskip 0pt
\begin{subfigure}{.18\textwidth}
  \centering
  \includegraphics[width=0.95\linewidth]{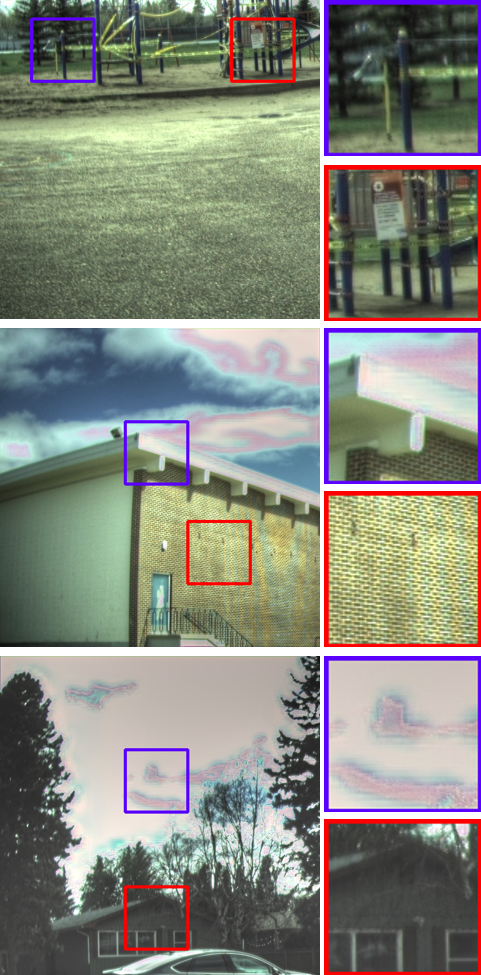}
  \caption{HDRCNN-Mask}
  \label{fig:supp_acmtog_hdrcnnmask2}
\end{subfigure}%
\begin{subfigure}{.18\textwidth}
  \centering
  \includegraphics[width=0.95\linewidth]{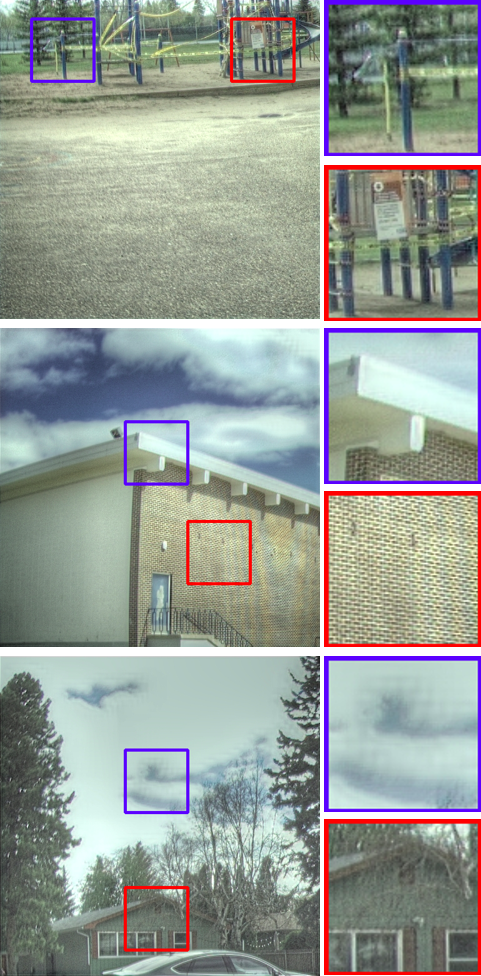}
  \caption{DPHR}
  \label{fig:supp_acmtog_dphr2}
\end{subfigure}%
\begin{subfigure}{.18\textwidth}
  \centering
  \includegraphics[width=0.95\linewidth]{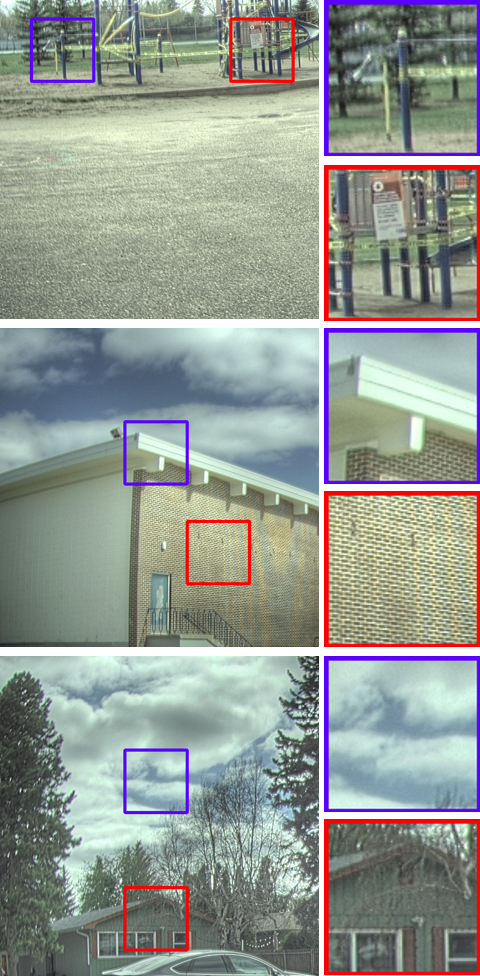}
  \caption{Ground truth}
  \label{fig:supp_acmtog_gt2}
\end{subfigure}%
\caption{Additional qualitative comparative results with state-of-the-art methods. The proposed DPHR method presents color consistency, free of visible artifacts, and overall able to recover richer textures in all poorly exposed regions. This is because DPHR is designed to handle polarization images, as it integrated the polarimetric cues into the framework. The Photomatix Enhanced TMO is used.}
\label{fig:supp_acmtog_soa2}
\end{figure*}

\begin{table*}[t!]
\caption{Quantitative comparative results with state-of-the-art methods (the higher the better except for MSE)}
\begin{center}
\label{tab:chapt5_DSSHDR_soa}
\begin{tabular}{c|c|c|c|c|c}
\hline
\small{Input sources} & \small{Methods} & \small{PU-PSNR} & \small{PU-SSIM} & \small{HDR-VDP2} &\small{MSE}\\ 
\hline
\multirow{2}{*}{Multiple-shot} & \small{HDR-GAN \cite{niu2020hdrgan}} & 15.62 $\pm$ 1.92 & 0.48 $\pm$ 0.018 & 50.54 $\pm$ 4.34 & 0.0599 $\pm$ 0.0037\\
& \small{DeepHDR \cite{wu2018deep}} & 18.69 $\pm$ 2.04 & 0.60 $\pm$ 0.037 & 49.55 $\pm$ 5.08 & 0.0293 $\pm$ 0.0043\\
\hline
\multirow{8}{*}{Single-shot} & \small{ENet \cite{Marnerides}} & 16.86 $\pm$ 1.45 & 0.88 $\pm$ 0.049 & 49.45 $\pm$ 3.61 & 0.0429 $\pm$ 0.0029\\
& \small{DrTMO \cite{Endo}} & 21.51 $\pm$ 1.74 & 0.84 $\pm$ 0.043 & 53.28 $\pm$ 3.28 & 0.0591 $\pm$ 0.0027\\ 
& \small{DRCP \cite{Liu}} & 16.48 $\pm$ 1.41 & 0.88 $\pm$ 0.063 & 49.25 $\pm$ 4.64 & 0.0784 $\pm$ 0.0062\\
& \small{HDRCNN \cite{Eilertsen}} & 21.82 $\pm$ 1.90 & 0.90 $\pm$ 0.066 & 53.66 $\pm$ 4.77 & 0.0391 $\pm$ 0.0037\\ 
& \small{HDRCNN-Mask \cite{Santos}} & 17.87 $\pm$ 1.39 & 0.83 $\pm$ 0.059 & 47.72 $\pm$ 3.22 & 0.0470 $\pm$ 0.0025\\
& \small{PHDR \cite{xuesong}} & 16.43 $\pm$ 1.91 & 0.73 $\pm$ 0.067 & 45.92 $\pm$ 4.45 & 0.0639 $\pm$ 0.0048\\
& \small{DPSHDR \cite{ting2021}} & 19.22 $\pm$ 2.13 & 0.83 $\pm$ 0.041 & 50.22 $\pm$ 4.36 & 0.0787 $\pm$ 0.0054\\
\cline{2-6}
& \small{\textbf{DPHR}} & \textbf{30.59 $\pm$ 3.19} & \textbf{0.94 $\pm$ 0.068} & \textbf{56.15 $\pm$ 5.28} & \textbf{0.0081 $\pm$ 0.0006}\\
\hline
\end{tabular}
\end{center}
\end{table*}

\subsection{Evaluation of the Training Loss Function}
In this experiment, we analyze the effect of each component of our training loss function. The results in Table~\ref{tab:chapt5_DSSHDR_ablation_loss} and Fig.~\ref{fig:chapt5_DSSHDR_ablation_loss} shows that $\mathcal{L}_r + \mathcal{L}_p$ loss is more powerful for preserving details as discussed in \cite{Santos}. We thus train our model using $\mathcal{L}_r + \mathcal{L}_p$ loss.

\subsection{Comparison with State-of-the-art Methods}
We use the test dataset to evaluate the proposed DPHR method and compare it with state-of-the-art methods, including seven deep-learning-based methods developed for the conventional cameras \cite{niu2020hdrgan, wu2018deep, Marnerides, Endo, Liu, Eilertsen, Santos}, one traditional model based method developed for the polarization camera \cite{xuesong}, and one deep-learning-based method developed also for the polarization camera \cite{ting2021}. For all methods, we used the codes and pretrain weights provided by the authors. Fig.~\ref{fig:chapt5_DSSHDR_soa}, \ref{fig:supp_acmtog_soa2} illustrates the qualitative results of the comparative study. We can observe that images reconstructed with the DPHR method can recover more accurate details in all properly exposed and poorly exposed regions, and alleviate color distortions and visible artifacts. The HDR-GAN \cite{niu2020hdrgan} and DeepHDR \cite{wu2018deep} results are relatively bright, and has difficulty restoring details in the over-exposed areas. The ENet \cite{Marnerides} result tends to be overly bright and smooth, as it over-enhances the extracted illumination features. Additionally, it's unable to improve HDR estimation for properly exposed regions. The DrTMO \cite{Endo} method suffers from blocking artifacts, and has difficulty recovering details in all properly exposed and poorly exposed regions. The DRCP \cite{Liu} method can restore details in the poorly exposed regions, but exhibits color distortions. Furthermore, the result is generally darker, thus contents lost in the under-exposed regions can not be restored. The PHDR \cite{xuesong} result is overall bright where pixels remain lost in the poorly exposed regions. The DPSHDR \cite{ting2021} result reveals more details, but remains unable to reconstruct sufficient details in both under- and over-exposed regions. 

The HDRCNN \cite{Eilertsen} method can recover some contents in the over-exposed sky region, but the result is overall dim and presents visual artifacts. The HDRCNN-Mask \cite{Santos} method follows the approach by HDRCNN to estimate details in over-exposed regions, and then reconstruct the final HDR image by combining with the input. However, unlike HDRCNN, HDRCNN-Mask proposes a feature masking mechanism to propagate valid features for properly exposed pixels. The output of HDRCNN-Mask is overly smooth with color artifacts, and reconstructs limited details of the cloud in the over-exposed sky area. Also, the content remains lost in the under-exposed areas. On the other hand, the proposed DPHR method presents color consistency, free of visible artifacts, and overall able to recover richer textures in both under- and over-exposed regions. This is because the proposed DPHR method is designed to handle polarization images, as we have integrated the polarization information (i.e., DoLP) into the network during training and inference to formulate the final HDR image. In addition to visual evaluation, the quantitative results are summarized in Table~\ref{tab:chapt5_DSSHDR_soa}. Since the proposed DPHR method uses DoLP as a strong accurate prior to help HDR recovery for polarization images, it performs favorably compared to state-of-the-art methods under various evaluation metrics, including both HDR (HDR-VDP2, MSE) and HDR tone-mapped LDR evaluation metrics (PSNR, SSIM, FSIM).

\begin{figure*}[t!]
\begin{subfigure}{0.7\textwidth}
  \centering
  \includegraphics[width=0.95\linewidth]{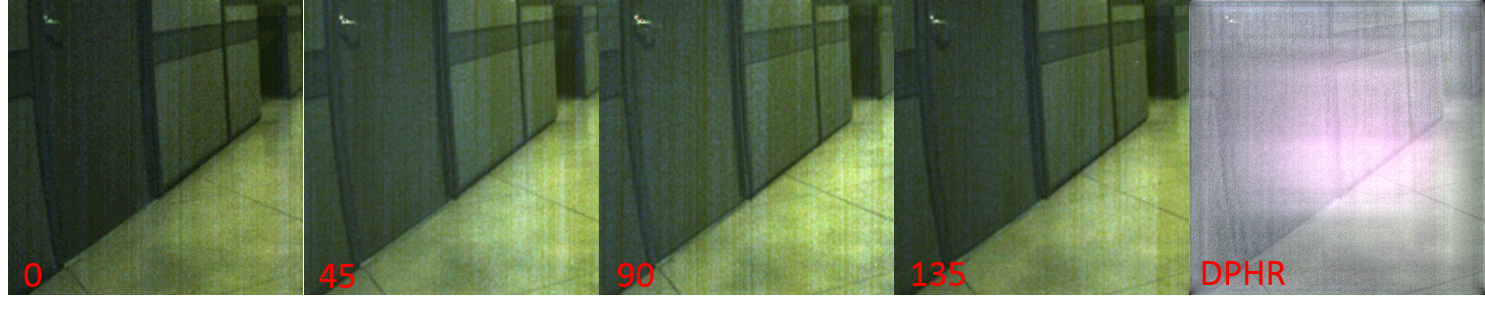}
  \caption{Low-light}
  \label{fig:DSSHDR_limitations_lowlight}
\end{subfigure}%
\centering
\vskip 0pt
\begin{subfigure}{0.7\textwidth}
  \centering
  \includegraphics[width=0.95\linewidth]{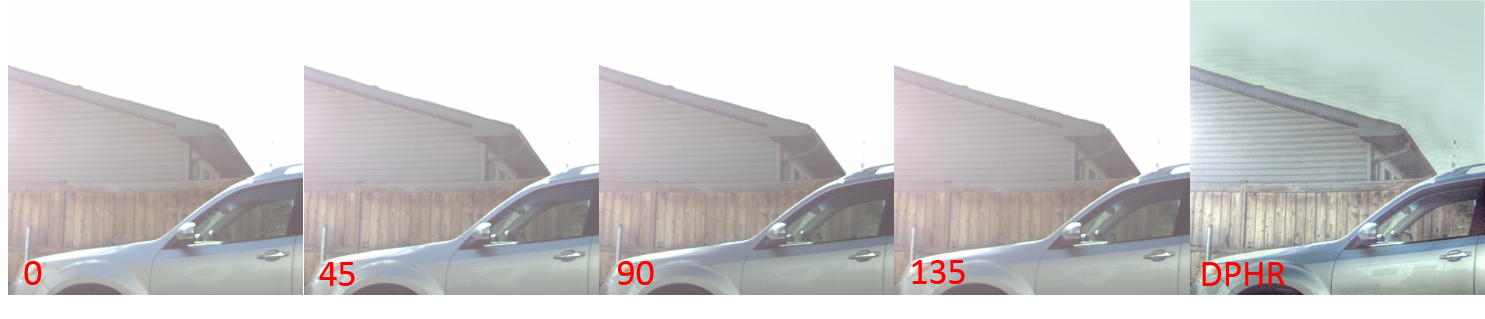}
  \caption{High dynamic-range}
  \label{fig:DSSHDR_limitations_highdr}
\end{subfigure}%
\caption{Limitation analysis. In the top row, the proposed DPHR method fails to reconstruct the indoor image captured from an poorly-lit setting. In the bottom row, the proposed DPHR method also fails to reconstruct the high dynamic-range image captured from an outdoor scene, where the details of the clouds in the sky are not recovered. The Photomatix Enhanced TMO is used.}
\label{fig:DSSHDR_limitations}
\end{figure*}


\section{Limitations and Future Work}
Despite that our method can reconstruct better details compared to existing works, its expressive power is still limited. Fig.~\ref{fig:DSSHDR_limitations} shows two examples of difficult scenes where our approach fails to reconstruct. In Fig.~\ref{fig:DSSHDR_limitations_lowlight}, our approach is unable to reconstruct the details and incorrectly reconstructs the color in the underexposed image captured from an indoor poorly-lit setting. This is mainly because our method is unable to handle noisy pixels. In particular, shot noise is prevalent when insufficient light reaches the sensor due to an exposure that is too brief for the lighting condition, and it results in random pixel variations which we observe as noisy pixels. This issue can be alleviated by exploring techniques to reduce the pixel noise in the pre-processing step. In Fig.~\ref{fig:DSSHDR_limitations_highdr}, our approach fails to recover the details of the clouds in the overexposed sky regions from an outdoor high dynamic-range scene. This is mainly due to images with broad regions of overexposed sunshine are scarce in our training dataset. This issue can be mitigated by augmenting our training dataset to incorporate such extreme cases to improve performance.

To follow up on this study, we will enrich our training dataset to improve the performance, we can collect a greater variety of scenes to include other outdoor scenes, such as buildings and people. Additionally, synthesizing HDR images using 3D computer graphics represents a promising avenue we would like to explore. Another interesting avenue of future work is to evaluate the benefits of the polarization camera with other image capturing devices \cite{Han2020neuromorphic, Bardallo2016AsyncCamera, Zhao2015ModuloCamera, Cull2015SnapshotCamera, Nayar, Suda, Rouf2011Glare} in reconstructing HDR images. We also plan to expand our method to other camera response functions, investigate demosaicing artifacts, and explore the use of the AoLP image, as an additional polarimetric cue, to further the image dynamic range and HDR detail recovery.


\section{Conclusion} \label{sec:Conclusion}
In this paper, we have proposed a novel learning based HDR reconstruction method using a polarization camera. The proposed DPHR method explores polarimetric information obtained by a polarization camera to enhance the reconstruction in all properly exposed and poorly exposed regions, which is a long-standing challenge in computer vision and robotics applications. In the proposed DPHR method, we first compute the input and mask image from the polarization images captured by the polarization camera, and use those cues in the feature masking mechanism to better propagate the valid features through the CNN. Then we combined the deep-learning predicted HDR image with a traditional model based HDR image to formulate the final HDR result. Our experimental results on the polarization image dataset show that the DPHR method demonstrated better qualitative and quantitative performances over state-of-the-art methods.

\bibliographystyle{IEEEtran}
\bibliography{main-tip}

\begin{thebibliography}{10}
\providecommand{\url}[1]{#1}
\csname url@samestyle\endcsname
\providecommand{\newblock}{\relax}
\providecommand{\bibinfo}[2]{#2}
\providecommand{\BIBentrySTDinterwordspacing}{\spaceskip=0pt\relax}
\providecommand{\BIBentryALTinterwordstretchfactor}{4}
\providecommand{\BIBentryALTinterwordspacing}{\spaceskip=\fontdimen2\font plus
\BIBentryALTinterwordstretchfactor\fontdimen3\font minus
  \fontdimen4\font\relax}
\providecommand{\BIBforeignlanguage}[2]{{%
\expandafter\ifx\csname l@#1\endcsname\relax
\typeout{** WARNING: IEEEtran.bst: No hyphenation pattern has been}%
\typeout{** loaded for the language `#1'. Using the pattern for}%
\typeout{** the default language instead.}%
\else
\language=\csname l@#1\endcsname
\fi
#2}}
\providecommand{\BIBdecl}{\relax}
\BIBdecl

\bibitem{Shakeri_2016_IROS}
M.~{Shakeri} and H.~{Zhang}, ``Illumination invariant representation of natural
  images for visual place recognition,'' pp. 466--472, 2016.

\bibitem{Shakeri_2017_ICCV}
------, ``Moving object detection in time-lapse or motion trigger image
  sequences using low-rank and invariant sparse decomposition,'' Oct 2017.

\bibitem{Shakeri_2019_CVPR}
------, ``Moving object detection under discontinuous change in illumination
  using tensor low-rank and invariant sparse decomposition,'' June 2019.

\bibitem{Yan_AttnHDR}
Q.~{Yan}, D.~{Gong}, Q.~{Shi}, A.~{van den Hengel}, C.~{Shen}, I.~{Reid}, and
  Y.~{Zhang}, ``Attention-guided network for ghost-free high dynamic range
  imaging,'' in \emph{2019 IEEE/CVF Conference on Computer Vision and Pattern
  Recognition (CVPR)}, 2019, pp. 1751--1760.

\bibitem{Lee_ExpBlendHDR}
S.~{Lee}, H.~{Chung}, and N.~I. {Cho}, ``Exposure-structure blending network
  for high dynamic range imaging of dynamic scenes,'' \emph{IEEE Access},
  vol.~8, pp. 117\,428--117\,438, 2020.

\bibitem{Marnerides}
D.~Marnerides, T.~Bashford-Rogers, J.~Hatchett, and K.~Debattista, ``Expandnet:
  A deep convolutional neural network for high dynamic range expansion from low
  dynamic range content,'' \emph{Computer Graphics Forum}, vol.~37, pp. 37--49,
  2018.

\bibitem{Eilertsen}
\BIBentryALTinterwordspacing
G.~Eilertsen, J.~Kronander, G.~Denes, R.~K. Mantiuk, and J.~Unger, ``Hdr image
  reconstruction from a single exposure using deep cnns,'' \emph{ACM
  Transactions on Graphics}, vol.~36, no.~6, p. 1–15, Nov 2017. [Online].
  Available: \url{http://dx.doi.org/10.1145/3130800.3130816}
\BIBentrySTDinterwordspacing

\bibitem{Santos}
\BIBentryALTinterwordspacing
M.~S. Santos, T.~I. Ren, and N.~K. Kalantari, ``Single image hdr reconstruction
  using a cnn with masked features and perceptual loss,'' \emph{ACM
  Transactions on Graphics}, vol.~39, no.~4, Jul. 2020. [Online]. Available:
  \url{https://doi.org/10.1145/3386569.3392403}
\BIBentrySTDinterwordspacing

\bibitem{Endo}
E.~Yuki, K.~Yoshihiro, and M.~Jun, ``Deep reverse tone mapping,'' \emph{ACM
  Transactions on Graphics}, vol.~36, no.~6, Nov. 2017.

\bibitem{Lee_2018a}
S.~{Lee}, G.~H. {An}, and S.~{Kang}, ``Deep chain hdri: Reconstructing a high
  dynamic range image from a single low dynamic range image,'' \emph{IEEE
  Access}, vol.~6, pp. 49\,913--49\,924, 2018.

\bibitem{Liu}
Y.~L. {Liu}, W.~S. {Lai}, Y.~S. {Chen}, Y.~L. {Kao}, M.~H. {Yang}, Y.~Y.
  {Chuang}, and J.~B. {Huang}, ``Single-image hdr reconstruction by learning to
  reverse the camera pipeline,'' in \emph{2020 IEEE/CVF Conference on Computer
  Vision and Pattern Recognition (CVPR)}, 2020, pp. 1648--1657.

\bibitem{xuesong}
X.~{Wu}, H.~{Zhang}, X.~{Hu}, M.~{Shakeri}, C.~{Fan}, and J.~{Ting}, ``Hdr
  reconstruction based on the polarization camera,'' \emph{IEEE Robotics and
  Automation Letters}, vol.~5, no.~4, pp. 5113--5119, 2020.

\bibitem{ting2021}
J.~{Ting}, X.~{Wu}, K.~{Hu}, and H.~{Zhang}, ``Deep snapshot hdr reconstruction
  based on the polarization camera,'' 2021.

\bibitem{Sen12}
P.~Sen, N.~K. Kalantari, M.~Yaesoubi, S.~Darabi, D.~B. Goldman, and
  E.~Shechtman, ``{Robust Patch-Based HDR Reconstruction of Dynamic Scenes},''
  \emph{ACM Transactions on Graphics}, vol.~31, no.~6, pp. 203:1--203:11, 2012.

\bibitem{Hu13}
J.~{Hu}, O.~{Gallo}, K.~{Pulli}, and X.~{Sun}, ``Hdr deghosting: How to deal
  with saturation?'' in \emph{2013 IEEE Conference on Computer Vision and
  Pattern Recognition}, 2013, pp. 1163--1170.

\bibitem{Oh15}
T.~{Oh}, J.~{Lee}, Y.~{Tai}, and I.~S. {Kweon}, ``Robust high dynamic range
  imaging by rank minimization,'' \emph{IEEE Transactions on Pattern Analysis
  and Machine Intelligence}, vol.~37, no.~6, pp. 1219--1232, 2015.

\bibitem{Lee14}
C.~{Lee}, Y.~{Li}, and V.~{Monga}, ``Ghost-free high dynamic range imaging via
  rank minimization,'' \emph{IEEE Signal Processing Letters}, vol.~21, no.~9,
  pp. 1045--1049, 2014.

\bibitem{Li14}
Z.~{Li}, J.~{Zheng}, Z.~{Zhu}, and S.~{Wu}, ``Selectively detail-enhanced
  fusion of differently exposed images with moving objects,'' \emph{IEEE
  Transactions on Image Processing}, vol.~23, no.~10, pp. 4372--4382, 2014.

\bibitem{Khademi_2017}
N.~K. Kalantari and R.~Ramamoorthi, ``Deep high dynamic range imaging of
  dynamic scenes,'' \emph{ACM Transactions on Graphics}, vol.~36, no.~4, 2017.

\bibitem{wu2018deep}
S.~Wu, J.~Xu, Y.-W. Tai, and C.-K. Tang, ``Deep high dynamic range imaging with
  large foreground motions,'' 2018.

\bibitem{liu2021adnet}
Z.~Liu, W.~Lin, X.~Li, Q.~Rao, T.~Jiang, M.~Han, H.~Fan, J.~Sun, and S.~Liu,
  ``Adnet: Attention-guided deformable convolutional network for high dynamic
  range imaging,'' in \emph{CVPRW}, 2021, pp. 463--470.

\bibitem{Pu_2020_ACCV}
Z.~Pu, P.~Guo, M.~S. Asif, and Z.~Ma, ``Robust high dynamic range (hdr) imaging
  with complex motion and parallax,'' in \emph{Proceedings of the Asian
  Conference on Computer Vision (ACCV)}, November 2020.

\bibitem{ma_mef}
K.~{Ma}, Z.~{Duanmu}, H.~{Zhu}, Y.~{Fang}, and Z.~{Wang}, ``Deep guided
  learning for fast multi-exposure image fusion,'' vol.~29, 2020, pp.
  2808--2819.

\bibitem{niu2020hdrgan}
Y.~Niu, J.~Wu, W.~Liu, W.~Guo, and R.~W.~H. Lau, ``Hdr-gan: Hdr image
  reconstruction from multi-exposed ldr images with large motions,'' 2020.

\bibitem{Kalantari2019deepvideo}
N.~Khademi~Kalantari and R.~Ramamoorthi, ``Deep hdr video from sequences with
  alternating exposures,'' \emph{Computer Graphics Forum}, vol.~38, pp.
  193--205, 05 2019.

\bibitem{AFR07}
A.~O. Akyüz, R.~Fleming, B.~Riecke, E.~Reinhard, and H.~Bülthoff, ``Do hdr
  displays support ldr content? a psychophysical evaluation,'' \emph{ACM
  Transactions on Graphics}, vol.~26, p.~38, 2007.

\bibitem{RTS07}
A.~Rempel, M.~Trentacoste, H.~Seetzen, H.~Young, W.~Heidrich, L.~Whitehead, and
  G.~Ward, ``Ldr2hdr: On-the-fly reverse tone mapping of legacy video and
  photographs,'' \emph{ACM Transactions on Graphics}, vol.~26, p.~39, 08 2007.

\bibitem{MSG17}
B.~Masia and D.~Gutiérrez, ``Dynamic range expansion based on image
  statistics,'' \emph{Multimedia Tools and Applications}, vol.~76, 01 2017.

\bibitem{KO14}
R.~P. {Kovaleski} and M.~M. {Oliveira}, ``High-quality reverse tone mapping for
  a wide range of exposures,'' in \emph{2014 27th SIBGRAPI Conference on
  Graphics, Patterns and Images}, 2014, pp. 49--56.

\bibitem{Masia16}
\BIBentryALTinterwordspacing
B.~Masia and D.~Gutiérrez, ``Content-aware reverse tone mapping,'' in
  \emph{Proceedings of the 2016 International Conference on Artificial
  Intelligence: Technologies and Applications}, 2016, pp. 235--238. [Online].
  Available: \url{https://doi.org/10.2991/icaita-16.2016.58}
\BIBentrySTDinterwordspacing

\bibitem{banterle2006inverse}
F.~Banterle, P.~Ledda, K.~Debattista, and A.~Chalmers, ``Inverse tone
  mapping,'' 01 2006, pp. 349--356.

\bibitem{kasliwal2015tonemap}
H.~Kasliwal and S.~Modi, ``A novel method for tone mapping of hdr images,''
  2015.

\bibitem{chen2021hdrunet}
X.~Chen, Y.~Liu, Z.~Zhang, Y.~Qiao, and C.~Dong, ``Hdrunet: Single image hdr
  reconstruction with denoising and dequantization,'' 2021.

\bibitem{Zhang2021DeepHE}
Y.~Zhang and T.~O. Aydin, ``Deep hdr estimation with generative detail
  reconstruction,'' \emph{Computer Graphics Forum}, vol.~40, 2021.

\bibitem{Han2020neuromorphic}
J.~Han, C.~Zhou, P.~Duan, Y.~Tang, C.~Xu, C.~Xu, T.~Huang, and B.~Shi,
  ``Neuromorphic camera guided high dynamic range imaging,'' in \emph{2020
  IEEE/CVF Conference on Computer Vision and Pattern Recognition (CVPR)}, 2020,
  pp. 1727--1736.

\bibitem{Bardallo2016AsyncCamera}
\BIBentryALTinterwordspacing
J.~A. Le\~{n}ero Bardallo, R.~Carmona-Gal\'{a}n, and
  A.~Rodr\'{\i}guez-V\'{a}zquez, ``Hdr image sensor with linear response and
  asynchronous detection of saturation: Demo,'' in \emph{Proceedings of the
  10th International Conference on Distributed Smart Camera}, ser. ICDSC
  '16.\hskip 1em plus 0.5em minus 0.4em\relax New York, NY, USA: Association
  for Computing Machinery, 2016, p. 204–205. [Online]. Available:
  \url{https://doi.org/10.1145/2967413.2974030}
\BIBentrySTDinterwordspacing

\bibitem{Zhao2015ModuloCamera}
H.~Zhao, B.~Shi, C.~Fernandez-Cull, S.-K. Yeung, and R.~Raskar, ``Unbounded
  high dynamic range photography using a modulo camera,'' in \emph{2015 IEEE
  International Conference on Computational Photography (ICCP)}, 2015, pp.
  1--10.

\bibitem{Cull2015SnapshotCamera}
C.~Fernandez-Cull, H.~Zhao, B.~Shi, B.~Tyrrell, J.~Lin, and R.~Raskar,
  ``Snapshot on-chip hdr roic architectures,'' 01 2015, p. CM3E.3.

\bibitem{Nayar}
S.~K. {Nayar} and T.~{Mitsunaga}, ``High dynamic range imaging: spatially
  varying pixel exposures,'' in \emph{Proceedings IEEE Conference on Computer
  Vision and Pattern Recognition (CVPR)}, vol.~1, 2000, pp. 472--479 vol.1.

\bibitem{Suda}
S.~Takeru, M.~T., M.~Yusuke, and O.~Masatoshi, ``Deep snapshot hdr imaging
  using multi-exposure color filter array,'' in \emph{Proceedings of the Asian
  Conference on Computer Vision (ACCV)}, 2020.

\bibitem{Rouf2011Glare}
M.~Rouf, R.~Mantiuk, W.~Heidrich, M.~Trentacoste, and C.~Lau, ``Glare encoding
  of high dynamic range images,'' in \emph{CVPR 2011}, 2011, pp. 289--296.

\bibitem{metzler2019deep}
C.~A. Metzler, H.~Ikoma, Y.~Peng, and G.~Wetzstein, ``Deep optics for
  single-shot high-dynamic-range imaging,'' 2019.

\bibitem{Sun_2020_LearnedOpticHDR}
Q.~Sun, E.~Tseng, Q.~Fu, W.~Heidrich, and F.~Heide, ``Learning rank-1
  diffractive optics for single-shot high dynamic range imaging,'' in \emph{The
  IEEE Conference on Computer Vision and Pattern Recognition (CVPR)}, June
  2020.

\bibitem{Martel2020NeuralSensor}
J.~N.~P. Martel, L.~K. Müller, S.~J. Carey, P.~Dudek, and G.~Wetzstein,
  ``Neural sensors: Learning pixel exposures for hdr imaging and video
  compressive sensing with programmable sensors,'' \emph{IEEE Transactions on
  Pattern Analysis and Machine Intelligence}, vol.~42, no.~7, pp. 1642--1653,
  2020.

\bibitem{Alghamdi2021Snapshot}
M.~Alghamdi, Q.~Fu, A.~Thabet, and W.~Heidrich, ``Transfer deep learning for
  reconfigurable snapshot hdr imaging using coded masks,'' \emph{Computer
  Graphics Forum}, vol.~40, 03 2021.

\bibitem{Schulz2019CodeMask}
------, ``{Reconfigurable Snapshot HDR Imaging Using Coded Masks and Inception
  Network},'' in \emph{Vision, Modeling and Visualization}, H.-J. Schulz,
  M.~Teschner, and M.~Wimmer, Eds.\hskip 1em plus 0.5em minus 0.4em\relax The
  Eurographics Association, 2019.

\bibitem{Fotiadou2020Snapshot}
K.~Fotiadou, G.~Tsagkatakis, and P.~Tsakalides, ``Snapshot high dynamic range
  imaging via sparse representations and feature learning,'' \emph{IEEE
  Transactions on Multimedia}, vol.~22, no.~3, pp. 688--703, 2020.

\bibitem{Serrano2016Snapshot}
\BIBentryALTinterwordspacing
A.~Serrano, F.~Heide, D.~Gutierrez, G.~Wetzstein, and B.~Masia, ``Convolutional
  sparse coding for high dynamic range imaging,'' \emph{Computer Graphics
  Forum}, vol.~35, no.~2, p. 153–163, May 2016. [Online]. Available:
  \url{http://dx.doi.org/10.1111/cgf.12819}
\BIBentrySTDinterwordspacing

\bibitem{Stokes}
D.~H. {Goldstein}, ``Polarized light,'' \emph{CRC press}, 2017.

\bibitem{Huynh_2010}
C.~P. {Huynh}, A.~{Robles-Kelly}, and E.~{Hancock}, ``Shape and refractive
  index recovery from single-view polarisation images,'' in \emph{in
  Proceedings of the IEEE Computer Society Conference on Computer Vision and
  Pattern Recognition (CVPR)}, 2010, pp. 1229--1236.

\bibitem{Debevec}
P.~E. Debevec and J.~Malik, ``Recovering high dynamic range radiance maps from
  photographs,'' in \emph{Proceedings of the 24th Annual Conference on Computer
  Graphics and Interactive Techniques}, 1997, p. 369–378.

\bibitem{Gatys_2016_CVPR}
L.~A. Gatys, A.~S. Ecker, and M.~Bethge, ``Image style transfer using
  convolutional neural networks,'' in \emph{Proceedings of the IEEE Conference
  on Computer Vision and Pattern Recognition (CVPR)}, June 2016.

\bibitem{Polcam}
``Imx250 cmos sensor,''
  \url{https://www.sony-semicon.co.jp/e/products/IS/industry/technology/polarization.html},
  2021.

\bibitem{EM_algo}
T.~K. {Moon}, ``The expectation-maximization algorithm,'' \emph{IEEE Signal
  Processing Magazine}, vol.~13, no.~6, pp. 47--60, 1996.

\bibitem{ronneberger2015unet}
R.~Olaf, F.~Philipp, and B.~Thomas, ``U-net: Convolutional networks for
  biomedical image segmentation,'' in \emph{Medical Image Computing and
  Computer-Assisted Intervention (MICCAI)}, ser. LNCS, vol. 9351.\hskip 1em
  plus 0.5em minus 0.4em\relax Springer, 2015, pp. 234--241.

\bibitem{Adam}
P.~K. Diederik and B.~Jimmy, ``Adam: A method for stochastic optimization,''
  \emph{International Conference on Learning Representations (ICLR)}, 2015.

\bibitem{hdrvdp2}
R.~Mantiuk, K.~J. Kim, A.~G. Rempel, and W.~Heidrich, ``Hdr-vdp-2: A calibrated
  visual metric for visibility and quality predictions in all luminance
  conditions,'' \emph{ACM Transactions on Graphics}, vol.~30, no.~4, Jul. 2011.

\bibitem{Mantiuk2021PU21}
R.~Mantiuk and M.~Azimi, ``Pu21: A novel perceptually uniform encoding for
  adapting existing quality metrics for hdr,'' 06 2021, pp. 1--5.

\end{thebibliography}

\end{document}